\documentclass[final,5p,twocolumn,times]{elsarticle}
\usepackage[utf8]{inputenc}
\usepackage{subfig}
\usepackage[numbers]{natbib}
\biboptions{sort&compress}
\usepackage{setspace}
\usepackage{amsmath}
\usepackage{nccmath}
\usepackage{amsfonts}
\usepackage{amsthm}
\usepackage{amssymb}
\usepackage{relsize}

\usepackage{multirow}
\usepackage{stackengine}
\newcommand\barbelow[1]{\stackunder[1.2pt]{$#1$}{\rule{.8ex}{.075ex}}}
\usepackage{float}
\usepackage{float}
\usepackage{ifthen}
\usepackage{etoolbox}
\usepackage{soul}
\usepackage[binary-units=true]{siunitx}
\usepackage{color}
\usepackage[off]{auto-pst-pdf}
\usepackage{array}
\usepackage{psfrag}
\usepackage{epstopdf}
\usepackage{epsfig}
\usepackage[prefix]{nomencl}

\usepackage{lipsum}
\usepackage{graphicx}
\usepackage{pgf}

\usepackage{pifont}
\usepackage{wasysym}

\usepackage[]{algorithmicx}
\usepackage{algpseudocode}
\usepackage{algorithm}
\usepackage{booktabs}

\algnewcommand{\LeftComment}[1]{\Statex \(\triangleright\) #1}

\usepackage[symbol]{footmisc}

\usepackage{titlesec}

\titleformat{\paragraph}
{\normalsize\normalfont\itshape}{\theparagraph}{1em}{}
\titlespacing*{\paragraph}
{0pt}{3.25ex plus 1ex minus .2ex}{1.5ex plus .2ex}

\usepackage[version=3]{mhchem} 
\usepackage[utf8]{inputenc}
\usepackage{datetime}
\usepackage{textcomp}
\usepackage{longtable}
\usepackage{amssymb}
\usepackage{cleveref}
\usepackage{tabularx}
\usepackage{url}
\usepackage{tikz}
\usetikzlibrary{calc,shapes,decorations,decorations.text, mindmap,shadings,patterns,matrix,arrows,intersections,automata,backgrounds, arrows.meta}
\usepackage[american,siunitx]{circuitikz}

\usetikzlibrary{positioning}
\usepackage{scalefnt}

\usetikzlibrary{arrows.meta,decorations.pathreplacing,positioning,shadows}

\usepackage{enumitem}
\setlist{nosep,leftmargin=*}

\usepackage[american,siunitx]{circuitikz}

\usepackage{pgfplots}
\usepackage{rotating}

\usepackage{rotating}
\usepackage{verbatim}
\usepackage{mwe}

\usepackage{framed} 
\usepackage{multicol} 
\usepackage{nomencl} 
\makenomenclature
\renewcommand*\nompreamble{\begin{multicols}{2}}
\renewcommand*\nompostamble{\end{multicols}}

\usepackage{etoolbox}
\renewcommand\nomgroup[1]{%
  \item[\itshape
  \ifstrequal{#1}{A}{Sets}{%
  \ifstrequal{#1}{B}{Variables}{%
  \ifstrequal{#1}{C}{Parameters}{%
  \ifstrequal{#1}{D}{Abbreviations}{
  \ifstrequal{#1}{E}{Tariff Nomenclature}{}}}}}%
]}

\allowdisplaybreaks
\journal{Renewable and Sustainable Energy Reviews}
\pgfplotsset{compat=1.14}

\begin{document}

\begin{frontmatter}



\title{Energy management of small-scale PV-battery systems: A systematic review considering practical implementation, computational requirements, quality of input data and battery degradation}


\author[label1]{Donald Azuatalam\corref{cor1}}
\address[label1]{School of Electrical and Information Engineering, The University of Sydney, Sydney, Australia}

\cortext[cor1]{Corresponding author}


\ead{donald.azuatalam@sydney.edu.au}

\author[label5]{Kaveh Paridari}
\address[label5]{Department of Electric Power and Energy Systems, KTH Royal Institute of Technology, Stockholm, Sweden}
\ead{paridari@kth.se}

\author[label1]{Yiju Ma}
\ead{yiju.ma@sydney.edu.au}

\author[label6]{Markus F\"{o}rstl}
\address[label6]{Institute for Electrical Energy Storage Technology, Technical University of Munich, Munich, Germany}
\ead{markus.foerstl@tum.de}


\author[label1]{Archie C. Chapman}
\ead{archie.chapman@sydney.edu.au}

\author[label1]{Gregor Verbi\v{c}}
\ead{gregor.verbic@sydney.edu.au}

\begin{abstract}
The home energy management problem has many different facets, including economic viability, data uncertainty and quality of strategy employed. The existing literature in this area focuses on individual aspects of this problem without a detailed, holistic analysis of the results with regards to practicality in implementation. In this paper, we fill this gap by performing a comprehensive comparison of seven 
different energy management strategies, each with different levels of practicality, sophistication and computational requirements. 
We analyse the results in the context of these three characteristics, and also critique the modelling assumptions made by each strategy. 
Our analysis finds that using a more sophisticated  energy  management strategy may not necessarily improve the performance and economic viability of the PV-battery system due to the effects of modeling assumptions, such as the treatment of uncertainties in the input data and battery degradation effects.
\end{abstract}



\end{frontmatter}


\begin{table*}[!t] \label{nomen}
   \begin{framed}
     \printnomenclature
   \end{framed}
\end{table*}

\nomenclature[A]{$\mathcal{D}$}{Set of days, $d \in \mathcal{D}$ in a year, $\mathcal{D} = \{1,...,365\}$}
\nomenclature[A]{$\mathcal{H}$}{Set of half-hour time-slots, $h \in \mathcal{H}$ in a day, \\ $\mathcal{H} = \{1,...,48\}$}
\nomenclature[A]{$\mathcal{C}$}{Set of customers, $c \in \mathcal{C}$}

\nomenclature[B]{$p^\mathrm{g+/-}$}{Power flowing from/to grid}
\nomenclature[B]{$p^\mathrm{b+/-}$}{Battery charge/discharge power}
\nomenclature[B]{$e^\mathrm{b}$}{Battery state of charge}
\nomenclature[B]{$s^\mathrm{b}$}{Battery charging status (0: discharge, 1: charge)}
\nomenclature[B]{$d^\mathrm{g}$}{direction of grid power flow (0: demand to grid, 1: grid to demand)}

\nomenclature[C]{$\eta^\mathrm{b+/-}$}{Battery charging/discharging efficiency}
\nomenclature[C]{$\bar{p}^\mathrm{b+/-}$}{Maximum battery charge/discharge power}
\nomenclature[C]{$\eta^\mathrm{b+/-}$}{Battery charging/discharging efficiency}
\nomenclature[C]{$\bar{e}^\mathrm{b}$}{Battery maximum state of charge}
\nomenclature[C]{$\barbelow{e}^\mathrm{b}$}{Battery minimum state of charge}
\nomenclature[C]{$p^\mathrm{pv}$}{Power from solar PV}
\nomenclature[C]{$\Delta h$}{Half hourly time steps}
\nomenclature[C]{$p^\mathrm{d}$}{Total customer demand}
\nomenclature[C]{$p^\mathrm{res}$}{Net demand}
\nomenclature[C]{$\bar{p}^\mathrm{g}$}{Maximum power taken from/to grid}

\nomenclature[D]{PV}{Photovoltaic}
\nomenclature[D]{DER}{Distributed energy resources}
\nomenclature[D]{FiT}{Feed in tariff}
\nomenclature[D]{ToU}{Time of use}
\nomenclature[D]{ToUA}{Time of use arbitrage}
\nomenclature[D]{SCM}{Self consumption maximisation}
\nomenclature[D]{BESS}{Battery energy storage system} 
\nomenclature[D]{IRR}{Internal rate of return}
\nomenclature[D]{PFA}{Policy function approximation}
\nomenclature[D]{MILP}{Mixed integer linear programming}
\nomenclature[D]{HEMS}{Home energy management system}
\nomenclature[D]{ANN}{Artificial neural network}
\nomenclature[D]{RNN}{Recurrent neural network}
\nomenclature[D]{SOC}{State of charge}
\nomenclature[D]{SOH}{State of health}
\nomenclature[D]{PFA}{Policy function approximation}
\nomenclature[D]{SCR}{Self consumption rate}

\nomenclature[E]{$T^\mathrm{tou}$}{Time-of-use energy charge} 
\nomenclature[E]{$T^\mathrm{fit}$}{Feed-in-tariff (FiT)}



\section{Introduction}
\label{intro}
\par Behind-the-meter distributed energy resources (DER), particularly solar PV and battery systems, are increasing in number in Australia and other parts of the world. As part of this, many energy management solutions employing DER have been proposed with the aim of minimising consumer's electricity costs as well as improving home comfort levels. Given the significant capital cost of these resources, home owners also seek to maximise their financial return on investment, and recover costs within a reasonable time frame. The economic benefits attainable therewith depend to a considerable extent on the performance of the energy management strategy deployed, and the tariff faced by the end user. 
From a practical point of view, energy management strategies can provide good quality and accurate solutions to the extent to which demand and PV forecasts can be accurately predicted while taking into account the computational limitations of a particular method. 
Our intent in this paper is to assess how various strategies fare in terms of their quality, practicality, sophistication and computational requirements, and to explain these outcomes in terms of the strategies employed and the modelling assumptions made when employing a given strategy.
In the following sections, we provide some relevant background information on this topic and motivations for the research and then we outline the main contributions of this paper.

\subsection{Background and motivation} \label{background}
\par The advancement in smart grid technologies has enabled a seamless integration of behind-the-meter distributed energy resources (DER) into distribution networks, where customers now have more control of their electricity use than ever before~\cite{jin2017foresee,kadavil2018application}. With supportive policies and the falling cost of these resources, we expect to see DER contributing significantly to the global energy mix in the near future. In Australia, for example, the price of residential solar PV systems dropped from roughly 12.5 \$/W in 2006 to about 2.42 \$/W in 2016 and projections by the Australian Energy Market Operator (AEMO) see an annual cost decline of 1.5\% for all PV system sizes until 2040. Furthermore, the prices for battery storage systems are predicted to fall from approximately 650 \$/kWh in 2017 to around 300 \$/kWh in 2037~\cite{aemosmall,ieapvps}. These massive drops in prices have caused a spike in the uptake of DER in Australia and other parts of the world~\cite{climatefc,climatecc}. 

\par In the residential setting, DER are typically managed by a \textit{home energy management system} (HEMS). The HEMS monitors and controls DER, and facilitates customer participation in demand response (DR) programs. 
Smart home energy management systems have been used in the recent literature to demonstrate how residential customers can effectively maximise their benefits for participating in DR schemes~\cite{keerthisinghe2018energy,keerthisinghe2016fast,jin2017foresee,kadavil2018application,jin2017user}. 

In more detail, HEMS are automated decision making tools for optimally scheduling DER and household appliances in order to minimise electricity cost as well as maintain an acceptable thermal comfort level. 
To fully benefit from a HEMS, electricity prosumers need to deploy an energy management strategy that results in higher electricity cost savings and recovers investment expenditures in a timely manner, given a particular retail tariff. This, however, depends on the availability and accuracy of day-ahead predictions of solar PV generation and demand. Since the load of a household is highly unpredictable, a simple energy management strategy will provide solutions based on the information available at the time of PV and/or battery installation. This implies the use of \textit{generic} load profiles, which are as close as possible to a consumer's typical load usage data (as a demand estimate), given various demographic (number and age of occupants) and home appliance details (electric or gas heating and cooking appliances), and then taking advantage of weather forecasts to predict PV generation. 
Although, this results a sub-optimal solution when compared to a scenario where more accurate demand profiles are known beforehand, or where accurate forecasts are possible, it gives a reasonable estimate of the power exchanged with the grid and consequently the overall cost of electricity. 
As time progresses, the energy management strategy can provide more accurate results based on observed load and PV usage patterns of the particular customer in question. 
This, in turn should result in greater cost savings for the end-user.

\par 
The HEMS strategies employed by residential customers with DER are affected by the existing tariff regime.
In several countries with high PV uptake rates, residential feed-in tariffs have been reduced at the same time as electricity retail prices have risen~\cite{IEA-energytaxes-2018}. 
In this context, \textit{self-consumption maximisation} (SCM) of the home solar PV generated power has become a commonly used energy management strategy; and currently, many smart inverters are capable of maximising the self-consumption of solar PV generated as a built-in feature. This essentially sets the PV-battery controller in the HEMS to self-consume as much solar power as possible and store the excess power in the battery to be used later in the evening. Here, the battery is only charged with power from the solar PV; and for maximum self-consumption, it is typical to charge the battery with the maximum possible charging rate and at the earliest time possible~\cite{struth2013pv}. 

An alternative to SCM is the use of \textit{Time of Use Arbitrage} (ToUA). This strategy involves pre-charging the battery with grid power during cheap off-peak periods, for use during periods of high electricity prices, when a customer faces ToU tariffs. This offers some extra security for power when there is a forecast of low PV generation. 
Both SCM and ToUA are heuristic approaches to energy management, and they do not explicitly minimise electricity cost. They do so by either storing excess PV generation and use it later to maximise self-consumption or by shifting consumption to low-price periods.
 
\par In addition to these two rule-based heuristic strategies, this work also examines the economic benefits of using principled energy management strategies based on optimisation methods, namely: 
\begin{itemize}
\item \textit{mixed integer-linear programming} (MILP), which explicitly minimises cost in an optimisation platform; 
\item \textit{dynamic programming} (DP), where the sequential decision energy management problem is cast as Markov decision processes (MDP), and 
\item \textit{policy function approximations} (PFA), which are neural networks trained with the output of one of the two optimisation-based HEMS approaches to provide fast online solutions, where in this work we use MILP\footnote{The choice of MILP is not essential to the PFA strategy; any other principled optimisation approach can be used to generate the training data as well.} to generate the training data. 
\end{itemize}
These home energy management strategies are analysed and compared in the context of assumptions made, available information and their level of practicality.

\subsection{Contributions}
In this work, we complete a comparative techno-economic assessment of different HEMS strategies considering forecast and computational limitations. Recent research in this area, assessing the techno-economic feasibility of PV-battery systems~\cite{merei2016optimization,linssen2017techno,quoilin2016quantifying,camilo2017economic,truong2016economics,schopfer2018economic,uddin2017techno,van2018techno,nicholls2015financial,hoppmann2014economic,nottrott2013energy} has neglected a detailed holistic comparison of several energy management strategies, such as the effects of forecast errors and modeling assumptions of the performance of a strategy.

Although a review of HEMS modelling and complexity was performed in~\cite{beaudin2015home}, the authors did not perform any simulation to support the findings of the review, and the state-of-the-art optimisation-based approaches were not considered in view of practical implementation. 
In light of this, in our previous work~\cite{azuatalam2018techno} we compared a principled optimisation technique, MILP, with three rule-based heuristic strategies.
Our results showed that well-tuned heuristic strategies can produce near-optimal solutions with lesser computational burden. 
However, several real-world features of the HEMS problem were overlooked, as where more-sophisticated HEMS strategies, and accordingly, the work reported in this paper extends our previous work to consider these facets of the problem.

Given this, the analysis in this paper extends the work done in~\cite{beaudin2015home} and preliminary results in our earlier conference paper~\cite{azuatalam2018techno} in the following ways: 

\begin{itemize}
	\item We perform a detailed literature review of energy management strategies in different contexts that affect their performance. 
	\item We model and implement seven different energy management strategies and evaluate their techno-economic performance, considering forecast uncertainties, computational limitations and battery degradation.
	\item We validate the performance of the energy management strategies using real customer load traces from the Ausgrid \textit{Solar Home Electricity Data} and retail tariffs used in the Sydney region of Australia. 
	\item We perform a detailed analysis of the simulation results in the light of modelling assumptions made and with respect to practical implementation, such as linearised battery operation and battery degradation effects.
	Our findings indicate the best HEMS strategy for a customer often depends on the quality of the data and accuracy of load and PV forecasts available.
\end{itemize}


\subsection{Organisation}
The remainder of the paper is organised as follows:  Section II presents a comprehensive literature review of the topic. 
In Section III, we present  an  overview  of  the  techno-economic  assessment  framework.  
Following this, we describe the different home energy management strategies in Section IV. 
The economic and technical (battery degradation) analyses are discussed in detail in Sections V and VI, respectively. 
In Section VII, we present and analyse the results, and in Section VIII, we discuss the results and conclude the paper.

\section{Literature review}
Methods and models for implementing smart home energy management systems have recently received considerable attention. In this section, we review relevant research covering energy management strategies in the context of areas related to this work, namely: cost savings; load and PV generation forecasting; temporal resolution; modelling complexity and computational feasibility; retail tariff settings, and; battery degradation.

\subsection{Cost savings}
\par Self-consumption maximisation (SCM) is a simple but effective heuristic approach for reducing energy costs.
In research studies, SCM has been frequently used as an energy management strategy to fully harness the potential of distributed energy resources, and it is made more effective by incorporating storage into the system, or by employing demand side management measures, as demonstrated in~\cite{struth2013pv,velik2013renewable,lorenzi2016comparing,widen2014improved,moshovel2015analysis,luthander2015photovoltaic}.
In~\cite{velik2013renewable}, SCM was used to obtain an average financial gain per day of up to 30\% in summer. 
Likewise, in~\cite{luthander2015photovoltaic}, the self-consumption rate (SCR, defined as the ratio between self-consumed power and generated PV power) was shown to increase by 13-24\% with a storage size of 0.5-\SI{1}{kWh} per installed \si{kW} PV size. 

\par However, batteries are useful in increasing PV self-consumption up to some certain kWh limit in relation to a particular PV size, after which benefits become marginal according to studies in~\cite{widen2013evaluating,thygesen2014simulation,huang2011quantifying}. 
Looking deeper, the studies in~\cite{nyholm2016solar,castillo2011pv} find that the extent to which battery storage can increase a customer's SCR is dependent on the actual household daily or average annual electricity consumption. 
In particular,~\cite{castillo2011pv,luthander2015photovoltaic} show that batteries perform best in maximising PV self-consumption when their capacity matches the customer's daily consumption. 

In light of the above, in our work, we consider the case where all customers have PV-battery systems. 
We compare the savings achieved by using a PV-battery system with principled optimisation-based HEMS strategies to the cost savings from using a simple heuristic strategies like SCM and ToUA, given same input parameters and assumptions.

\subsection{Load and PV generation forecasting}
Forecasts of load and PV generation are essential inputs to any optimisation-based HEM strategy. 
While PV forecasting is relatively straight-forward with a good weather forecasts, forecasting individual customer's loads is quite challenging.
Nonetheless, many studies in this area assume perfect foresight of PV and demand, even though this is not possible in reality. 

Accurate day-ahead forecast of PV generation, which is dependent on precise weather forecasts, is integral to the maximisation of PV-battery SCR (self-consumption rate) and consequently on cost savings~\cite{zhang2013metrics,lazos2014optimisation}.  
In view of this,~\cite{petrollese2018use} investigated the impacts of weather forecast errors on the self consumption rate of PV-battery systems. 
Their results show an SCR reduction of 0.5-4.5\%, with global horizontal irradiance (GHI) root mean square error in the range of 0.622-\SI{1.687}{kWh/m^2/day}. Furthermore,~\cite{wang2015operational} implemented a receding-horizon optimisation to minimise the adverse effects arising from inaccurate PV generation forecasting. 
On the other hand,~\cite{lodl2011operation,marinelli2014testing} showed that battery storage systems can cushion the impacts resulting from solar PV forecast errors. 
Similarly, ~\cite{masa2014improving} show how to use batteries to minimise the difference in annual SCR due to forecast errors.

In contrast, customer day-ahead demand, is generally harder to predict accurately, because it depends on more uncertain factors when compared to PV output prediction~\cite{rodrigues2014daily,aguera2018weather}. 
One reasonable way to forecast a day's demand is to use load profile for the previous week, as consumers usually have a typical weekly consumption pattern~\cite{struth2013pv}. This is known as a \textit{persistence forecast}. 
This method  was implemented in~\cite{moshovel2015analysis}, and resulted in a 4.4\% reduction in SCR compared to a perfect forecast scenario. This SCR reduction was achieved, however, using a PI controller that minimises forecast uncertainties. 
In this work, we also apply a simple persistence forecast algorithm, but instead use it to analyse the impacts of forecast errors on optimisation-based energy management approaches and on a larger set of economic indicators.

\subsection{Temporal resolution}
Past studies in the area of PV-battery optimisation and self-consumption have used low resolution PV and demand data, typically of either 30 minutes~\cite{abdulla2016optimal,abdulla2017importance,ratnam2015scheduling,keerthisinghe2016fast,ratnam2015optimization,zhang2015optimal} or 1 hour~\cite{aghajani2015presenting,nugraha2015optimization,fuselli2013action,liu2015mpc,lorenzi2016comparing,lu2005short,luna2016optimal,ming2017multi,nunez2017optimal,pezeshki2014model,ranaweera2016optimization,shang2016generation,su2014stochastic,teng2013optimal,wu2014optimal,zhang2015optimal}. 
This is mainly due to the unavailability of data with finer resolution, since household smart meters are usually set to record data in 30 minutes intervals. 
However, demand and PV temporal resolution can have a significant impact on PV-battery self consumption~\cite{beck2016assessing,cao2014impact,tangimpacts,wyrsch2013effect,hawkes2005impacts,kools2016data}. 
In~\cite{beck2016assessing}, the impacts of the temporal resolution of demand and solar PV generation on PV-battery self consumption were studied. They concluded that adding storage to PV systems have the effect of mitigating the influence of lower resolution data on SCR.
A similar outcome is found in~\cite{wyrsch2013effect}. 
However, with PV systems alone, a 15 minute resolution data is required for PV and demand in order to achieve reasonably accurate results. 
For the aforementioned reasons, in this paper, we use half-hourly resolution data for PV-battery systems, in order to minimise SCR errors.

\subsection{Modelling complexity and computational feasibility}
Computational complexity of an optimisation problem is measured by its time and space computational cost. 
The complexity of any HEMS-solution method depends on a range of factors, including modelling simplicity and/or model assumptions, input or model description size, and the algorithmic complexity of the energy scheduling strategy deployed.

Several advanced optimisation-based approaches have been proposed for solving the home energy management problem. 
They include: principled optimisation approaches, such as
\textit{mixed-integer linear programming} (MILP)~\cite{MILP1,MILP2,KeerthisingheVerbicChapman_AUPEC2014,nottrott2013energy,hassan2017optimal,beck2016assessing,luna2017mixed,bozchalui2012optimal,hubert2011realizing}, 
\textit{stochastic optimisation}~\cite{yao2016residential,samadi2014real,sou2011scheduling}, 
\textit{robust optimisation}~\cite{du2016robust,chen2012real}, 
\textit{dynamic programming} (DP)~\cite{TischerVerbic2011,li2014optimal,jeddi2017dynamic,ranaweera2016optimization}, 
\textit{approximate dynamic programming} (ADP)~\cite{keerthisinghe2016fast}; 
local search and evolutionary methods, such as \textit{particle swarm optimisation} ~\cite{pedrasa2009improved,PSO2,pedrasa2010coordinated,pedrasa2011robust} and \textit{genetic algorithms} ~\cite{pena2017optimizing,miyazato2016multi} 
The problem is typically formulated in the spirit of model predictive control~\cite{jin2017foresee,zhu2015switched,tazvinga2014energy} although this might not be explicitly stated.

\par In general, the objective of a HEMS is to provide reasonably fast, accurate energy management schedules. To do this well, it should also consider uncertainties in input data and operating characteristics of the devices under its control~\cite{garifi2018stochastic}. 
For example, if the uncertainties in the load or PV output, and the nonlinearities in the battery and inverter operation are both considered, the model better captures reality, but at the expense of modelling simplicity and computational speed. 
Thus, there is a trade-off between model complexity and computational burden.

At one extreme, this can be seen in the case of DP, which results in high-quality solutions by incorporating both uncertainties and nonlinearities. 
On the other hand, the use of MILP, which approximates (linearises) the battery operating characteristics and lies on deterministic forecasts, results in a faster solution but with a lower quality. 
To address the computational challenge of DP, authors in~\cite{TischerVerbic2011} approximated the influence of uncertainties in PV and demand while authors in~\cite{KeerthisingheVerbicChapman_TSG2016} used ADP with temporal difference learning to obtain a faster solution with only a slight reduction in quality. Furthermore, even faster, near-optimal solutions can be obtained with \textit{policy function approximations} (PFA) using machine learning. 
This involves generating training data from a high-quality solver such as DP with customers' historic data, and training the neural networks on this data offline, then executing the the policy encoded in the PFA online.  This effectively moves the computational task to the offline phase of the HEMS method, thereby greatly reducing the online computation time. 

\par Robust optimisation (RO) methods have been adopted in~\cite{du2016robust,chen2012real}, to model the uncertainties in household energy usage, while in~\cite{yao2016residential,samadi2014real}, stochastic optimisation was used (SO).
Although, the use of DP and RO yields tractable results, these approaches are sensitive to the choice of the uncertainty set (or structure of the uncertainty)~\cite{bertsimas2011theory}. 
In~\cite{chen2012real}, the use of both stochastic and robust optimisation methods for modelling the uncertainties in household energy consumption were compared. Numerical results show that while RO solves the cost minimisation problem with lesser computational burden, the use of SO results in lower electricity bills for the consumer.  

\par With evolutionary methods and other local search methods, a main disadvantage is that they offer no guarantee with regards to optimality. 
However, they can return reasonably good solutions if uncertainties are catered for, as shown in~\cite{pedrasa2011robust}. 
The HEMS can also be operated based on simple rule-based heuristics~\cite{parra2016effect,linssen2017techno}. Some commonly employed strategies include \textit{self-consumption maximisation} and \textit{time-of-use arbitrage}. These have the advantage of simplicity and computational speed, but the solutions are sub-optimal in terms of cost minimisation when compared to principled optimisation approaches.

\subsection{Retail tariff}
The impacts of residential retail tariff and FiT on the performance of PV-battery systems was studied in~\cite{parra2016effect,ren2016modelling,hassan2017optimal}. 
In~\cite{parra2016effect}, authors showed that the use of flat tariffs is more profitable than using either ToU or dynamic (wholesale) tariffs with PV-battery systems that perform simple PV energy time-shift. 
On the contrary, in a study involving UK households~\cite{hassan2017optimal}, the cost savings incurred with wholesale electricity prices was only 5\% higher than that with ToU tariffs. 
However, results in~\cite{ren2016modelling} showed critical peak retail pricing (CPP) to be more profitable than flat tariffs with PV-battery systems in the Australian context. The profitability of using a particular retail tariff over the other depends on the actual tariff figures, the peak period (in the case of ToU and CPP), and customers response. Moreover, distribution network service companies (DNSPs) are allowed to adjust their prices yearly to account for varying revenue in order to approach a predetermined amount (revenue-cap), being revenue regulated entities. Therefore, customers' price response in one year, has an effect on tariff prices the following year, which in turn affects the financial return on customers with DER. In this research, however, we consider households on ToU tariffs only, because cost savings are maximised when PV-battery systems are operated with ToU tariffs.


\subsection{Battery degradation}
Battery aging and degradation is not often included in operational studies, but has been studied in~\cite{magnor2009concept,koller2013defining,naumann2017simses,truong2016economics,benjaminb}. 
In~\cite{koller2013defining}, a model predictive control-based battery energy storage system (BESS) control strategy with explicit modelling of battery degradation cost function was used for peak shaving purposes.
A simplified battery aging model for efficient BESS operation is presented in~\cite{benjaminb}. Their results show that the monetary value of the battery increases by 74\% when aging is considered. 
A heuristic float aging model based on \textit{Arrhenius’ law} is presented in~\cite{magnor2009concept} to predict Lithium-ion battery lifetime expectancy. However, authors assumed that the battery aging during the charging and discharging phases is the same. 

A more detailed battery degradation and simulation model (SimSES) is introduced in~\cite{naumann2017simses} to enhance the profitability of the BESS. This model has been applied in~\cite{truong2016economics} to study the techno-economic analysis of the PowerWall battery. 
The simulation results showed a battery capacity degradation of 36\% after the system 20-year lifespan. The SimSES tool, which mainly focuses on the techno-economic analysis of Lithium-ion batteries, is employed in this study due to its flexibility and accuracy, compared to other open-source battery simulation software.


\section{Methodology}
In this section, we describe the modelling framework and the system topology adopted in this research. Furthermore, we provide information on relevant data used in this study.
\subsection{Framework and system topology}
We adopt the framework in Figure~\ref{framework} to examine the techno-economic feasibility of PV-battery systems of residential customers considering both perfect and persistence forecast of solar PV generation and demand, using the above-mentioned energy management strategies. More specifically, we carry out financial analysis and battery degradation study on the different energy scheduling strategies. The economic viability indicators include the annual cost savings and the internal rate of return (IRR) of investment, while the battery degradation indicators include the battery state of health (SOH), annual full equivalent cycles (FEC), and expected lifetime (EBL).  
Demand, PV output, DER and tariff data are fed as input to the energy management scheduling algorithm. The output of this algorithm that includes the grid power, battery charge/discharge power, and the battery SOC is then used to carry out the financial analysis and battery degradation study.


\par Figure~\ref{topology_a} depicts the system topology. Here we have assumed that the PV and battery systems are connected to two separate inverters, but PV-battery systems with one common inverter for both the PV and battery as depicted in Figure~\ref{topology_b} are also common. The power from the PV, $P_\mathrm{pv}$, is either exported to the grid or feeds the load $P_\mathrm{d}$ directly. 
The battery power, $P_\mathrm{b}$, is bi-directional for charge/discharge. During the discharge phase, the battery power can directly supply the demand or is fed to the grid. The grid power, $P_\mathrm{g}$, which is also bi-directional, is used to charge the battery during the charging phase, or fed to the demand typically when PV and battery power is not available.

\tikzstyle{linepart} = [draw, very thick, -latex', dashed]
\tikzstyle{decision} = [diamond, draw, fill=blue!20, 
text width=4.5em, text badly centered, node distance=3cm, inner sep=0pt]
\tikzstyle{block3} = [rectangle, draw, text width=7.5em, text centered, rounded corners, very thick, minimum height=2em]
\tikzstyle{block2} = [rectangle, draw, text width=7.5em, text centered, rounded corners, very thick, minimum height=2em]
\tikzstyle{block} = [rectangle, draw, text width=8em, text centered, rounded corners, very thick, minimum height=3em]
\tikzstyle{line} = [draw, -latex, thick]
\tikzstyle{cloud} = [draw, ellipse,fill=red!20, node distance=3cm,
minimum height=1em]
\tikzstyle{block4} = [rectangle, draw, text width=9em, text centered, rounded corners, very thick, minimum height=2.2em]
\tikzstyle{block5} = [rectangle, draw, text width=10em, text centered, rounded corners, very thick, minimum height=2.2em]

\begin{figure}[t!]
	\footnotesize
	\begin{center}
		\begin{tikzpicture}[node distance = 2cm, auto]
		
		\node [block5] (input) {\textbf{Input Data} \\ Tariff data, DER data \\Load \& PV output data };
		
		\node [block5, below of=input, node distance=1.65cm] (energy) {\textbf{Energy Management \\ Approach} \\ Rule-based heuristics, Optimisation, PFA};
		
		\node [block4, below left = 0.5cm and 0.1cm  of energy, node distance=2cm] (finance) {Financial Analysis};
		\node [block4, below right = 0.5cm and 0.1cm  of energy, node distance=2cm] (degrad) {Degradation Study};
		
		\path[line] let \p1=(energy.south), \p2=(finance.north) in (energy.south) --  +(0,0.4*\y2-0.4*\y1) -| node [pos=0.3, above] {} (finance.north);
		\path[line] let \p1=(energy.south), \p2=(degrad.north) in (energy.south) -- +(0,0.4*\y2-0.4*\y1) -| node [pos=0.3, above] {} (degrad.north);
		
		
		\node [block4, below of=finance, node distance=1.2cm] (indices) {Annual savings, IRR};
		
		\node [block4, below of=degrad, node distance=1.2cm] (health) {SOH, FEC, EBL};
		
		%
		\path [line] (input) -- (energy);	
		\path [line] (degrad) -- (health);
		\path [line] (finance) -- (indices);
		
		
		
		

		
		\end{tikzpicture}
		\caption{Modelling framework.}
		\label{framework}
	\end{center}
\end{figure}
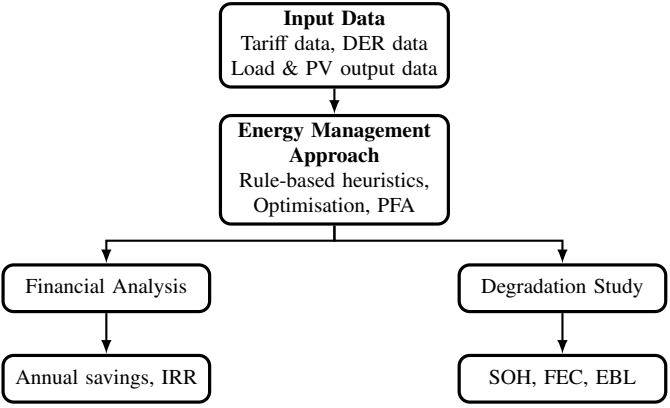

\tikzstyle{blocks} = [rectangle, draw, very thick, text width=1em, rounded corners, minimum height=1em]
\tikzstyle{lines} = [draw, latex-latex new]
\newcommand{\tarrow}[1]{#1 & \tikz[>=#1] \draw[<->] (1.8,0) -- (2.3,0);}

\pgfplotsset{compat=1.14}	
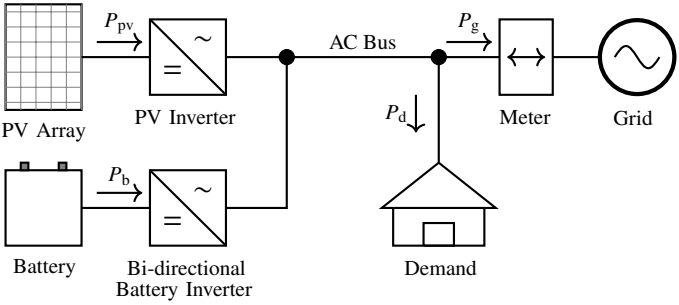
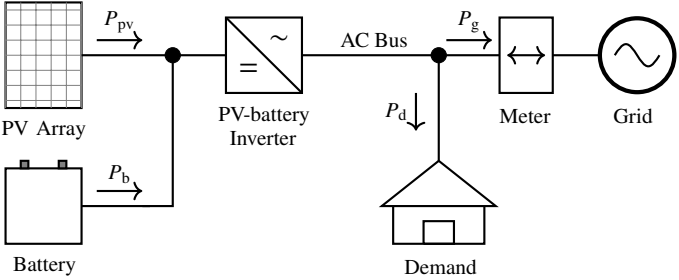
\begin{figure}[t!]
\footnotesize
\begin{tabular}{c}
\hspace{-1.4em}
\captionsetup[subfigure]{oneside,margin={-2cm,0cm}}
\subfloat[Separate PV and battery inverters]{
    	\begin{circuitikz}
    	\centering
			\draw[thick] (-5.2,-0.7) node[above]{} rectangle (-4.2,0.7);
			\draw[color=gray, help lines, line width=.05pt] (-5.2,-0.7)
			grid[xstep=.2cm, ystep=.2cm] (-4.2,0.7);
			\node[text width=3cm] at (-3.75,-1) {PV Array};
			\draw [thick,->] (-4,0.2) to node[above] {$P_\mathrm{pv}$} (-3.4,0.2);
			\draw[thick] (-4.2,0) -- (-3.3,0);
			
			\draw[thick] (-3.3,-0.5) rectangle (-2.3,0.5);
			\draw[thick] (-3.3,0.5) -- (-2.3,-0.5);
			\node at (-2.6,0.2) [very thick, scale = 1.5 ]{$\sim$};
			\node at (-3.0,-0.2) [very thick, scale = 1.5] {$=$};
			\node[text width=3cm] at (-2,-0.8) {PV Inverter};
			\draw[thick] (-2.3,0) -- node[above]{AC Bus} (1.3,0);

			\draw[preaction={fill=black}] (-1.5,0) node[above]{} circle (0.1);
			
			\draw[thick] (-5.2,-2.5) rectangle (-4.2,-1.5);
			\draw[thick,fill=gray] (-5,-1.5) rectangle (-4.9,-1.4);
			\draw[thick,fill=gray] (-4.5,-1.5) rectangle (-4.4,-1.4);
			\draw[thick] (-4.2,-2) -- (-3.3,-2);
			\node[text width=3cm] at (-3.6,-2.8) {Battery};
			
			\draw [thick,->] (-4,-1.8) to node[above] {$P_\mathrm{b}$} (-3.4,-1.8);
			\draw[thick] (-2.3,-2) -- (-1.5,-2) -- (-1.5,0);
			
			\draw[thick] (-3.3,-2.5) rectangle (-2.3,-1.5);
			\draw[thick] (-3.3,-1.5) -- (-2.3,-2.5);
			\node at (-2.6,-1.8) [very thick, scale = 1.5 ]{$\sim$};
			\node at (-3.0,-2.2) [very thick, scale = 1.5] {$=$};
			\node[text width=3cm] at (-2.1,-2.8) {Bi-directional};
			\node[text width=3cm] at (-2.25,-3.1) {Battery Inverter};
			
			
			\draw[preaction={fill=black}] (0.5,0) node[above]{} circle (0.1);
			\draw[thick] (0.5,0) -- (0.5,-1.4);
			\draw [thick,->] (0.2,-0.5) to node[left] {$P_\mathrm{d}$} (0.2,-1);
			\draw[thick] (-0.25,-2) -- (0.5,-1.4) -- (1.25,-2) -- cycle;
			\draw[thick] (-0.1,-2) -- (-0.1,-2.5) -- (1.1,-2.5) -- (1.1,-2);
			\draw[thick] (0.3,-2.5) -- (0.3,-2.2) -- (0.7,-2.2) -- (0.7,-2.5) -- (0.3,-2.5);
			\node[text width=3cm] at (1.55,-2.8) {Demand};
			
		    \draw [thick,->] (0.6,0.2) to node[above] {$P_\mathrm{g}$} (1.2,0.2);
		    \draw[thick] (1.3,-0.5) rectangle (2,0.5);
		    \draw[-latex, thick, <->] (1.4,0) -- (1.9,0) node[] {};
		    \node[text width=3cm] at (2.8,-0.8) {Meter};
		    \draw[thick] (2,0) -- (2.6,0);
		    \draw[thick] (3.6,0) node[oscillator]{};
		    \node[text width=3cm] at (4.3,-0.8) {Grid};
	\end{circuitikz} \label{topology_a} }  \vspace{0.5cm} \\
\hspace{-1.4em}
\captionsetup[subfigure]{oneside,margin={-2.3cm,0cm}}
\subfloat[Common PV-battery inverter]{
 			\begin{circuitikz}
 			\centering
			\draw[thick] (-5.2,-0.7) node[above]{} rectangle (-4.2,0.7);
			\draw[color=gray, help lines, line width=.05pt] (-5.2,-0.7)
			grid[xstep=.2cm, ystep=.2cm] (-4.2,0.7);
			\node[text width=3cm] at (-3.75,-1) {PV Array};
			\draw [thick,->] (-4,0.2) to node[above] {$P_\mathrm{pv}$} (-3.4,0.2);
			\draw[thick] (-4.2,0) -- (-2.3,0);
			
			\draw[thick] (-2.3,-0.5) rectangle (-1.3,0.5);
			\draw[thick] (-2.3,0.5) -- (-1.3,-0.5);
			\node at (-1.6,0.2) [very thick, scale = 1.5 ]{$\sim$};
			\node at (-2.0,-0.2) [very thick, scale = 1.5] {$=$};
			\node[text width=3cm] at (-0.9,-0.8) {PV-battery};
			\node[text width=3cm] at (-0.75,-1.1) {Inverter};
			\draw[thick] (-1.3,0) -- node[above]{AC Bus} (0.6,0);
			\draw[thick] (0.6,0) -- (1.3,0);			
			
			\draw[preaction={fill=black}] (-3.0,0) node[above]{} circle (0.1);
			
			\draw[thick] (-5.2,-2.5) rectangle (-4.2,-1.5);
			\draw[thick,fill=gray] (-5,-1.5) rectangle (-4.9,-1.4);
			\draw[thick,fill=gray] (-4.5,-1.5) rectangle (-4.4,-1.4);
			\draw[thick] (-4.2,-2) -- (-3.3,-2);
			\node[text width=3cm] at (-3.6,-2.8) {Battery};
			
			\draw [thick,->] (-4,-1.8) to node[above] {$P_\mathrm{b}$} (-3.4,-1.8);
			\draw[thick] (-3.4,-2) -- (-3.0,-2) -- (-3.0,0);
			
			
			
			\draw[preaction={fill=black}] (0.5,0) node[above]{} circle (0.1);
			\draw[thick] (0.5,0) -- (0.5,-1.4);
			\draw [thick,->] (0.2,-0.5) to node[left] {$P_\mathrm{d}$} (0.2,-1);
			\draw[thick] (-0.25,-2) -- (0.5,-1.4) -- (1.25,-2) -- cycle;
			\draw[thick] (-0.1,-2) -- (-0.1,-2.5) -- (1.1,-2.5) -- (1.1,-2);
			\draw[thick] (0.3,-2.5) -- (0.3,-2.2) -- (0.7,-2.2) -- (0.7,-2.5) -- (0.3,-2.5);
			\node[text width=3cm] at (1.55,-2.8) {Demand};
			
		    \draw [thick,->] (0.6,0.2) to node[above] {$P_\mathrm{g}$} (1.2,0.2);
		    \draw[thick] (1.3,-0.5) rectangle (2,0.5);
		    \draw[-latex, thick, <->] (1.4,0) -- (1.9,0) node[] {};
		    \node[text width=3cm] at (2.8,-0.8) {Meter};
		    \draw[thick] (2,0) -- (2.6,0);
		    \draw[thick] (3.6,0) node[oscillator]{};
		    \node[text width=3cm] at (4.3,-0.8) {Grid};
	\end{circuitikz} \label{topology_b}
} \\
\end{tabular}
\caption{PV-battery connection topologies used in the paper.}
\label{figure9}
\end{figure}

\subsection{Electricity retail tariffs}

\par
\begin{figure}[t]
	\centering
	\includegraphics[scale = 0.55]{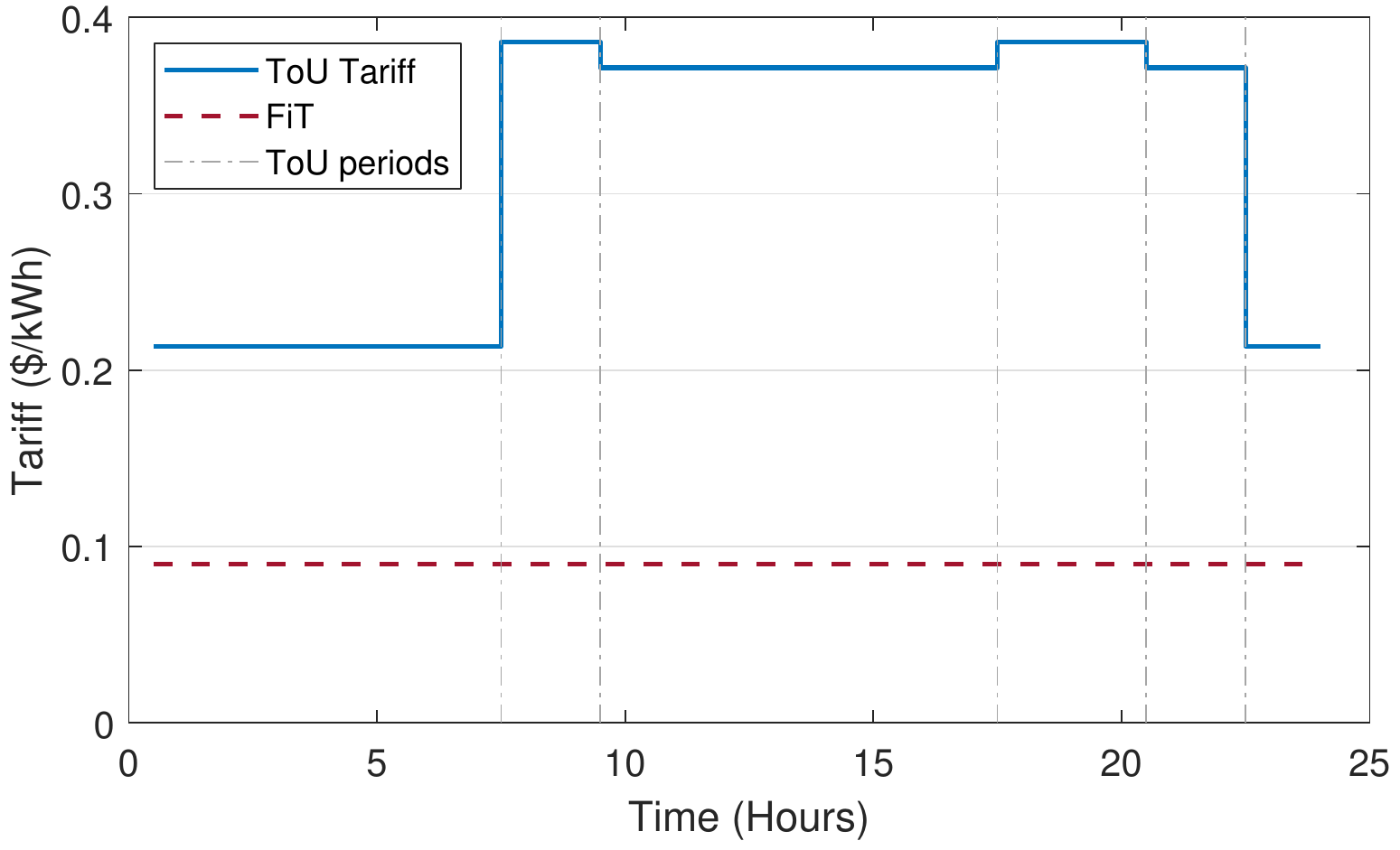}
	\caption{Retail (ToU) and Feed-in tariffs.}
	\label{figure3}
\end{figure}

 The retail charges used in this study are obtained from Origin Energy\footnote{Origin Energy NSW Residential Energy Price Fact Sheet for Essential Energy Distribution Zone. Available at https://www.originenergy.com.au \\ ToU tariff, Peak period:  7am to 9am, 5pm to 8pm; shoulder period: 9am to 5pm, 8pm to 10pm; off-peak period: 10pm to 7am.}, one of the big three electricity retailers in New South Wales, Australia. Table~\ref{table1} shows the residential electricity prices of Origin Energy for customers in the Essential energy distribution zone (depicted in Figure~\ref{figure3}, which also shows the ToU time-differentiated periods). These prices include the actual cost of electricity, retailer's risk management and service fees, and the network (Essential Energy) charge. 

\begin{table}[t]
	\footnotesize
	\centering
	\caption{Retail charges}
	\label{table1}
	\begin{tabular}[t]{c@{\hspace{0.12cm}}c@{\hspace{0.12cm}}c@{\hspace{0.12cm}}c@{\hspace{0.12cm}}c@{\hspace{0.12cm}}c@{\hspace{0.12cm}}c}
		\hline \hline
		\multicolumn{1}{l}{
			\begin{tabular}[c]{@{}c@{}}Tariff\\ Type\end{tabular}} & \begin{tabular}[c]{@{}c@{}}Fixed\\ charge\\ (\$/day)\end{tabular} & \begin{tabular}[c]{@{}c@{}}All\\ Usage\\ (c/kWh)\end{tabular} & \begin{tabular}[c]{@{}c@{}}Off peak\\ Usage\\ (c/kWh)\end{tabular} & \begin{tabular}[c]{@{}c@{}}Shoulder\\ Usage\\ (c/kWh)\end{tabular} & \begin{tabular}[c]{@{}c@{}}Peak\\ Usage\\ (c/kWh)\end{tabular} & \begin{tabular}[c]{@{}c@{}}Feed-in\\ Tariff\\ (c/kWh)\end{tabular} \\ \hline
		ToU & 1.551 & - & 21.340 & 37.147 & 38.588 & 9.0 \\ \hline
	\end{tabular}
\end{table}

\begin{table}[t]
	\footnotesize
	\centering
	\caption{PV and battery size combinations}
	\label{table2}
	\begin{tabular}{lcc}
		\hline \hline
		Solar PV size & Battery size & Battery Type \\
		(kWp) & (kWh) & \\
		\hline
		3 - 4 & 6.5 (LG Chem RESU 6.5) & Lithium ion\\ 
		5 - 6 & 9.8 (LG Chem RESU 10) & Lithium ion\\
		7 - 10  & 14 (Tesla Power Wall 2) & Lithium ion\\
		\hline
	\end{tabular}
\end{table}

\begin{table}[t]
	\footnotesize
	\centering
	\caption{Battery specifications}
	\label{table3}
	\begin{tabular}[t]{c@{\hspace{0.4cm}}c@{\hspace{0.2cm}}c@{\hspace{0.2cm}}c@{\hspace{0.2cm}}c@{\hspace{0.2cm}}}
		\hline \hline
		\multicolumn{1}{l}{
			\begin{tabular}[c]{@{}c@{}}Battery Type\\ \end{tabular}} & \begin{tabular}[c]{@{}c@{}}Nominal\\capacity\\ (kWh)\end{tabular} & \begin{tabular}[c]{@{}c@{}}Usable\\capacity\\ (kWh)\end{tabular} &  \begin{tabular}[c]{@{}c@{}}Max.\\Power\\ (kW)\end{tabular} &  \begin{tabular}[c]{@{}c@{}}Round-trip\\efficiency\\ (\%)\end{tabular} \\ \hline
		LG Chem RESU 6.5 & 6.5 & 5.9 & 4.2 & 95 \\ 
		LG Chem RESU 10 & 9.8 & 8.8 & 5.0 &  95 \\ 
		Tesla Power Wall 2 & 14.0 & 13.5 & 5.0 & 90 \\ \hline
	\end{tabular}
\end{table}



\begin{figure}[t]
	\centering
	\includegraphics[scale = 0.55]{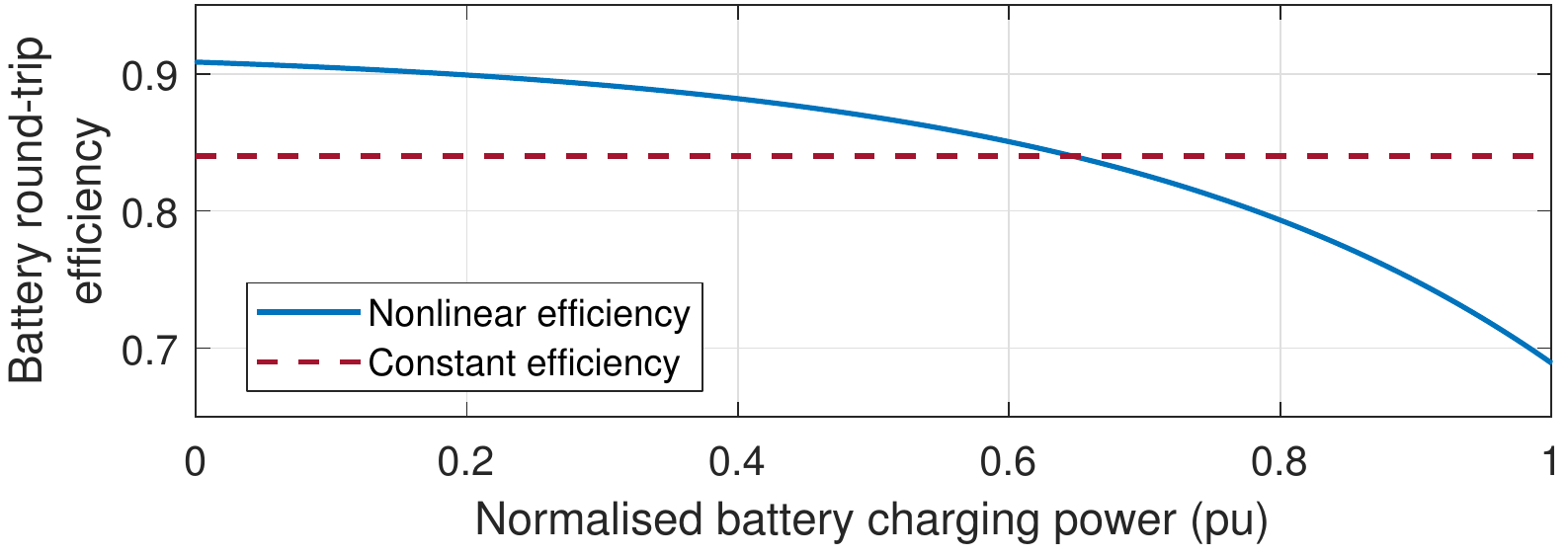}
	\hbox{\hspace{0.6em} \includegraphics[scale = 0.55]{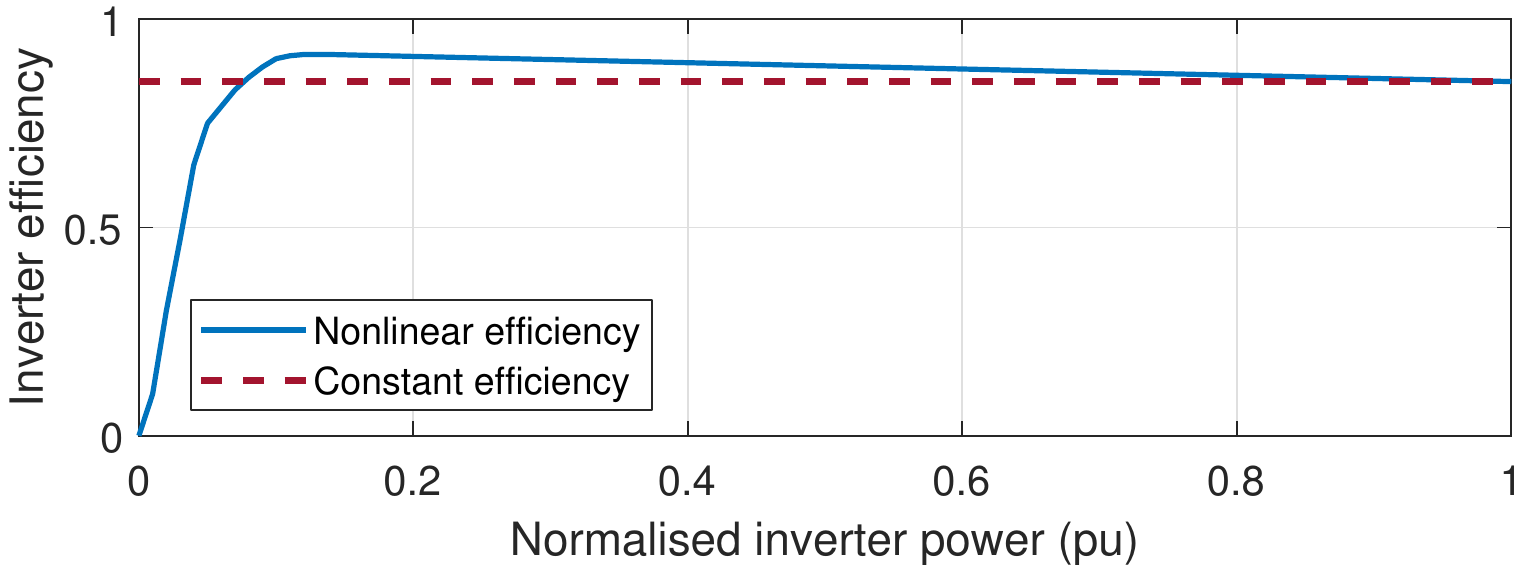}}
	\caption{Generic battery (top) and inverter (bottom) efficiency curves.}  
\label{figure4}
\end{figure}

\subsection{Load and DER data}

We sourced the demand and solar PV generation profiles from the Ausgrid (another DNSP in NSW) \textit{Solar Home Electricity Data}~\cite{solarhome}. This dataset comprise three years of half-hourly resolution electricity demand and solar PV data for the period between July 2010 to June 2013. Although the dataset contains anomalous or incomplete demand and PV data for some customers due to reasons like inverter failure, 52 clean customer load and PV profiles have been extracted by Ratnam et. al~\cite{ratnam2017residential} and are used in this paper.
The size of the solar PV ranges from 3 to 10 kWp (in steps of 1kW), with an average value of about 5 kW (which corresponds to the average PV size in Australia). Statistically, 9.60\%, 28.85\%, 38.5\%, 7.69\%, and 15.38\% of the 52 customers have installed PV capacities of 3.0, 4.0, 5.0,
6.0, and 7.0-10 kWp, respectively. The battery size of the customer depends on the size of the solar PV installed. In Australia, typically, 1.5-3 \si{kWh} of storage is used per 1 \si{kW} of PV installed ~\cite{aemosmall}. The PV-battery size combinations are shown in Table~\ref{table2} while the battery specifications are given in Table~\ref{table3}~\cite{aginnovators}. With reference to Figure~\ref{topology_a}, the PV inverter efficiency has already been accounted for in the dataset, so we have assumed a PV inverter efficiency of 1 in our simulations.

\subsection{Battery operating model} \label{batterymodel}

\par The battery round-trip efficiencies, given by the manufacturers (Table~\ref{table3}) do not include the battery inverter efficiency, and also they have not considered losses due to the energy drawn from the battery management system. 
SMA technologies have completed a comprehensive study in~\cite{smaefficiency} on the average efficiency of lithium-ion batteries. This study shows that the round-trip efficiency of a typical lithium-ion battery including losses is around 84\%. Furthermore, authors in~\cite{barnes2015semi,bennett2015development,braam2014peak,gitizadeh2013effects,hanna2014energy,hoevenaars2012implications,hoke2013look,hong2013interactive,koohi2014smart,kusakana2015daily,li2014optimal,hong2006short,olaszi2017comparison,ranaweera2016optimization,reddy2017optimal,ren2016optimal,schibuola2017influence,syed2016predictive,torreglosa2015energy,wang2014control,ranaweera2017distributed} have also assumed a constant efficiency for the battery. However, research done in ~\cite{pandvzic2019accurate,raszmann2017modeling} showed that an inaccurate battery model can lead to an overestimate of battery’s economic performance.

\par On the contrary, authors in~\cite{keerthisinghe2016fast,keerthisinghe2014evaluation,tischer2011towards} used nonlinear efficiency curves both for the battery and the inverter, since, in reality, efficiencies are not constant. For batteries, they depend on many factors like SOC, temperature, charge and discharge rates, etc. By contrast, the authors in~\cite{riffonneau2011optimal} use an explicit battery model based on a linear dependence of the battery voltage on the SOC, and a nonlinear inverter efficiency based on a quadratic interpolation of an experimental curve and a quadratic function representing the losses.

The inverter and battery efficiency curves used in~\cite{keerthisinghe2016fast}, are shown in Figure~\ref{figure4} for illustration. The nonlinear battery efficiency is generic and will vary depending on the chemistry. Nonetheless, the figure qualitatively shows that the battery efficiency is nonlinear and therefore, assuming a constant battery efficiency for the HEMS model can lead to overestimation of its economic performance. However, a detailed study done in~\cite{stevens1996study}, showed that on the average, the overall battery charging efficiency is around 91\%. This value is close to that obtained in~\cite{smaefficiency}. Therefore, we have used a constant battery charge/discharge efficiency value of 91\% (round-trip value of 84\%) for our simulation and analyses. This is also depicted in Figure~\ref{figure4}. Nonetheless, we show in our results (in Section~\ref{scheduling}) that this could result in infeasible results for energy management strategies using a linear battery model.

\section{Energy management models and simulation assumptions} 
In this section, we describe the different energy management strategies and the simulation assumptions made in this work. Due to the space constraints, we only provide a summary of the strategies considered, which is however detailed enough to make the paper sufficiently self-contained. More details can be found in the respective references.

\subsection{Home energy management strategies}	
The home energy management strategies considered in the paper are listed below, in an increasing level of sophistication.

\begin{itemize}
	\item Rule-based heuristic approaches 
	\begin{itemize}
		\item Self consumption maximisation 
		\item Time of use arbitrage 
		\item Self consumption maximisation with ToU arbitrage
	\end{itemize}
	\item Optimisation approaches
	\begin{itemize}
		\item Mixed integer linear programming
		\item Dynamic programming 
	\end{itemize}
	\item Policy function approximations (machine learning approaches)
	\begin{itemize}
		\item PFA with generic day types
		\item PFA with customer specific day types
	\end{itemize}
\end{itemize} 

\subsubsection{Self-consumption maximisation (SCM)}
Here, the inverter is set such that the energy from the PV first meets demand fully, with surplus energy used in charging the battery or fed back to the grid. This method works well for most consumers and its the baseline strategy employed by retailers and battery suppliers. The SCM algorithm is described in Algorithm~\ref{SCM Algorithm}\footnote{See Nomenclature for definition of variables and parameters used in Algorithms 1 and 2}. 

\begin{algorithm}[t]
	\caption{Self-consumption maximisation algorithm}\label{SCM Algorithm}

	\footnotesize
	$\mathcal{C}$: set of customers \\
	$\mathcal{D}$: set of days in a year\\
	$\mathcal{H}$: set of half hours in a day \\

	\begin{algorithmic}[1]
		\For{each customer $c \in \mathcal{C}$}
		\State \textbf{Read} yearly load and PV profile
		\State \textbf{Initialize} $p_{d,h}^\mathrm{g+}, p_{d,h}^\mathrm{g-}, p_{d,h}^\mathrm{b+}, p_{d,h}^\mathrm{b-}, p_{d,h}^\mathrm{res}, e_{d,h}^\mathrm{b}$ to zero vectors 
		\State Set $e_{1,1}^\mathrm{b}$ = $0.5\bar{e}^\mathrm{b}$
		\For {each day $d \in \mathcal{D},\ h \in \mathcal{H}$}
		\State $p_{d,h}^\mathrm{res} = p_{d,h}^\mathrm{pv} - p_{d,h}^\mathrm{d}$\Comment{First meet demand with PV}
		\If{$p_{d,h}^\mathrm{res} > 0$}\Comment{PV power greater than demand} 
			\State $e_{d,h}^\mathrm{b,res} = \bar{e}^\mathrm{b} - e_{d,h}^\mathrm{b}$
		\If{$e_{d,h}^\mathrm{b,res} > 0$}
				\State $p_{d,h}^\mathrm{b+} = \min(\bar{p}^\mathrm{b+},\min(\left|p_{d,h}^\mathrm{res}\right|,e_{d,h}^\mathrm{b,res}/\eta^\mathrm{b+}\Delta h))$\Comment{Charge Bat.}
				\State $p_{d,h}^\mathrm{g-} = \left|p_{d,h}^\mathrm{res}\right| - p_{d,h}^\mathrm{b+}$\Comment{Export remainder}
			\Else
                \State $p_{d,h}^\mathrm{g-} = \left|p_{d,h}^\mathrm{res}\right|$
			\EndIf
		\ElsIf{$p_{d,h}^\mathrm{res} < 0$}\Comment{PV power less than demand}
		  \State $e_{d,h}^\mathrm{b,res} = e_{d,h}^\mathrm{b} - \barbelow{e}^\mathrm{b}$
		  \If{$e_{d,h}^\mathrm{b} \approx \barbelow{e}^\mathrm{b}$}\Comment{Bat. is almost empty}
		  	\State $p_{d,h}^\mathrm{g+} = \left|p_{d,h}^\mathrm{res}\right|$ \Comment{Import}
		  \Else \Comment{Bat. is not empty}
		    \State $p_{d,h}^\mathrm{b-} = \min(\bar{p}^\mathrm{b-},\min(\left|p_{d,h}^\mathrm{res}\right|,e_{d,h}^\mathrm{b,res}\eta^\mathrm{b-}/\Delta h))$\Comment{Discharge Bat.}
		    \State $p_{d,h}^\mathrm{g-} = \left|p_{d,h}^\mathrm{res}\right| - p_{d,h}^\mathrm{b-}$\Comment{Import remainder}
		    \EndIf
		\Else \Comment{PV power equals demand} 
		  \State $p_{d,h}^\mathrm{g+/-} = 0,\ p_{d,h}^\mathrm{b+/-} = 0$
		\EndIf
		\State $e_{d,h+1}^\mathrm{b} = e_{d,h}^\mathrm{b} + \Delta h(\eta^\mathrm{b+}p_{d,h}^\mathrm{b+} - p_{d,h}^\mathrm{b-}/\eta^\mathrm{b-})$
		\EndFor
		\EndFor
	\end{algorithmic}
\end{algorithm}

\subsubsection{Time-of-use arbitrage (ToUA)}
This strategy is similar to self-consumption maximisation but involves pre-charging the battery to a certain pre-determined SOC using cheap off-grid power, to be used later during the day when electricity prices are higher. Intuitively, this method is only beneficial with ToU tariffs and for certain customers whose load profile is well suited.

\subsubsection{Self-consumption maximisation with time-of-use arbitrage (SCM+ToUA)}
This is a hybrid of the SCM and ToUA strategies. The baseline strategy here is SCM, but ToUA is applied only where there is a perfect forecast of low PV generation.

\subsubsection{Mixed integer linear programming (MILP)} This method, unlike the first two, explicitly takes the actual electricity cost and FiT into account in an optimisation framework. The objective of this HEMS approach~\eqref{milp_obj_fn} is to minimise electricity cost, given known fixed tariff prices over a decision horizon. In this work, we evaluate the deterministic version of MILP, which implies that the stochasticity of the energy management problem has been neglected and this results in a lower quality solution. The full details of this optimisation problem can be found in~\cite{azuatalam2017impacts}. It is worth mentioning that the battery SOC transition equation (in constraint 2) has been linearised.


\begin{align} \label{milp_obj_fn}	
\underset{\substack{p_{d,h}^\mathrm{g+}, p_{d,h}^\mathrm{g-}, p_{d,h}^\mathrm{b+}, p_{d,h}^\mathrm{b-}, \\ d_{d,h}^\mathrm{g}, s_{d,h}^\mathrm{b}, e_{d,h}^\mathrm{b}, }}{\text{minimise}}
& \sum\limits_{d \in \mathcal{D}} \bigg[ \sum\limits_{h \in \mathcal{H}}  T^\mathrm{tou} p_{d,h}^\mathrm{g+} - T^\mathrm{fit}p_{d,h}^\mathrm{g-} \bigg] 
\end{align}
subject to: 
\begin{enumerate}
\setlength\itemsep{0.05em}
 \item Power balance constraints
 \item Battery SOC constraints
 \item Maximum grid connection limits
 \item Upper and lower limits on continuous variables
\end{enumerate}

	\begin{algorithm}[t]
	\caption{MILP rolling horizon algorithm}\label{MILP optimisation Algorithm}
	\footnotesize
	
	$\mathcal{C}$: set of customers \\
	$\mathcal{D}$: set of days in a year\\
	$\mathcal{H}$: set of half hours in a day \\
	
	
	
	

	\begin{algorithmic}[1]
		\For{each customer $c \in \mathcal{C}$}
		\State Read yearly load and PV profile
		\State Set $e_{1,1}^\mathrm{soc}$ = $0.5\bar{e}^\mathrm{soc}$
		\For {each day $d \in \mathcal{D},\ h \in \mathcal{H}$}
		\State Solve~\eqref{milp_obj_fn} for day $d$ to $d+1$\Comment{two-day rolling horizon}
		\State \textbf{Return} $p_{d,h}^\mathrm{g+}$, $p_{d,h}^\mathrm{g-}$ $e_{d,h}^\mathrm{soc}$ for day $d$
		\State $\mathrm{Set}$ $e_{d+1,1}^\mathrm{soc}$ = $e_{d,\left\vert\mathcal{H}\right\vert}^\mathrm{soc}$
		\EndFor
		\EndFor
	\end{algorithmic}
\end{algorithm}

\subsubsection{Deterministic dynamic programming (DP)} In this strategy, the HEMS is formulated as a sequential stochastic optimisation problem, with the objective of minimising a cost function. Here, the variations in electrical demand and solar PV are incorporated in the optimisation process as non-controllable inputs with random variables. Also, the nonlinearities in the battery and inverter operation are captured in a nonlinear transition function that describes how the battery SOC evolves over time with respect to charging and discharging efficiency and the maximum charging rate. These consequently enables DP to give higher quality solutions compared to the other strategies.  

\par To solve the underlying HEMS problem using DP, we first model it as a Markov decision process (MDP). An MDP consists of a \textit{state space}, $(s\in \mathcal{S})$, a \textit{decision space}, $(x\in \mathcal{X})$,  \textit{transition functions} and \textit{contribution functions}. Let the index $k$ be a particular time-step and $K$ be the total number of time-steps. A state variable, $s_{k}\in \mathcal{S}$, contains the information that is necessary and sufficient to make the decisions and compute costs, rewards and transitions. The decision variable, $x_{k}\in \mathcal{X}$, is the control action for the battery for the decision horizon while the random variable, $\omega_{k}\in \mathrm{\Omega}$, represents exogenous information such as customer behavioural patterns and weather conditions~\cite{keerthisinghe2016energy}. Since this is the deterministic version of DP, we have neglected the effects of $\omega_{k}$. Therefore, the general form of the MDP problem is: 
\begin{align} \label{dp_obj_fn}
\underset{\pi}{\text{min}}\ & \mathbb{E}\left\{ \sum^{K}_{k=0} C^{\pi}_{k}({s}_{k},{ x}_{k},{\omega}_{k})\right\}
\end{align}
subject to: 
\begin{enumerate}
\setlength\itemsep{0.05em}
 \item Power balance constraints
 \item Battery SOC constraints
 \item Maximum grid connection limits
 \item Upper and lower limits on continuous variables
\end{enumerate}
\vspace{0.5em}
\noindent where: 
\begin{itemize}
 \item $\pi$ is a choice of action (policy) for each state, $\pi:\mathcal{S}\rightarrow \mathcal{X}$, which represents the charge/discharge status of the battery over the decision horizon.
 \item The contribution function $C_{k}({s}_{k},{x}_{k},{\omega}_{k})$ is the cost/reward of energy incurred at a given time-step $k$ and accumulates over time  \item The transition function ${{s}_{k+1}={s}^{M}\left({s}_{k},{x}_{k}\right)}$
describes the evolution of states from time step $k$ to next time step $k+1$. ${s}^{M}(.)$ represents the battery operating model. 

\end{itemize}

\par DP solves the optimisation problem (MDP) using \textit{value iteration}, by computing the expected future discounted cost (\textit{value function} $V^{\pi}({s}_{k})$) of following a policy, $\pi$, starting in state, ${s}_{k}$, and is given by:
\begin{align}\label{dp_eqn_a}
{\displaystyle V^{\pi}({s}_{k})= \sum_{{s}'\in \mathcal{S}}\mathbb{P}({s}'|{s}_{k},{x}_{k},{\omega}_{k})\left[C({s}_{k},{x}_{k},{s}')+ V^{\pi}({s}')\right]} \end{align} 


\noindent where $\mathbb{P}({s}'|{s}_{k},{x}_{k},{\omega}_{k})$ is the transition probability of landing on state ${s}'$ from ${s}_{k}$ if we take action ${x}_{k}$. The expression in~\eqref{dp_eqn_a} is a recursive reformulation of the objective function. To find the optimal solution or the optimal value function: $V_{k}^{\pi^{*}}({s}_{k})$, we need to solve~\eqref{dp_eqn_b} for each state, where $\pi^{*}$ is an optimal policy.

\begin{align}\label{dp_eqn_b} 
V_{k}^{\pi^{*}}({s}_{k})=\min\limits_{{x}_{k}\in \mathcal{X}_{k}} \left(C_{k}({s}_{k}, {x}_{k}({s}_{k})) + \mathbb{E}\left\{ V_{k+1}^{\pi^{*}}({s}')|{s}_{k}\right\} \right).
\end{align}

More details on the modelling of DP can be found in~\cite{keerthisinghe2016energy,KeerthisingheVerbicChapman_PSCC2014,tischer2011towards}.


\subsubsection{Policy function approximation using customer specific day types (PFAS)}
This policy function approximation algorithm is a non-parametric model that is first trained offline using historic data. The trained network is then used to make computationally fast solutions of the HEMS problem online. The underlying function that maps inputs to outputs is an artificial neural network (depicted in Figure~\ref{fig5}) trained using customer historic PV and demand data, and the output of the MILP algorithm (optimal policy), based on the Levenberg-Marquardt backpropagation training method. Thereafter, the trained network specific to each customer is used to predict the grid power for the current year. 

%
%
%
%
	
\tikzstyle{blocks} = [rectangle, draw, very thick, text width=7em, text centered, rounded corners, minimum height=9em]
\tikzstyle{lines} = [draw, -latex']
	
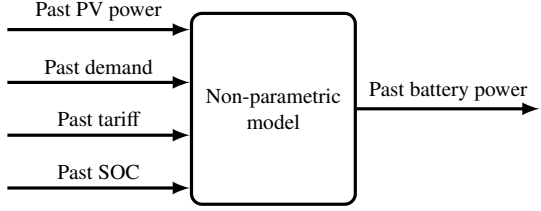
\begin{figure}[t]
		\centering
		\begin{circuitikz}
			\footnotesize
\node [blocks] (input) {Non-parametric model};
 \draw[-latex, black, very thick] (-3.5,1.05)--(-1.1,1.05) node[pos=0.5,above] {Past PV power};
  \draw[-latex, black, very thick] (-3.5,0.35)--(-1.1,0.35) node[pos=0.5,above] {Past demand};
   \draw[-latex, black, very thick] (-3.5,-0.35)--(-1.1,-0.35) node[pos=0.5,above] {Past tariff};
  \draw[-latex, black, very thick] (-3.5,-1.05)--(-1.1,-1.05) node[pos=0.5,above] {Past SOC};
    \draw[-latex, black, very thick] (1.1,0)--(3.5,0) node[pos=0.5,above] {Past battery power};
		\end{circuitikz}
			\caption{Policy function approximation - training phase}
	\label{fig5}
	\end{figure}
	
\par In the online (execution) phase, data for the current year is fed as input to the model. For time slot $h$, the non-parametric model is used to generate the battery SOC for time slot $h+1$ (calculated using the battery power for time slot $h$), which is fed back as for the next time slot. This time-series prediction continues until the end of the decision horizon.	


Since the constraints on the battery operation are not considered in the PFA formulation, the computed SOC might violate some of these constraints. Therefore, a control strategy is applied to ensure the constraints are adhered to. This strategy simply applies a filter on SOC$(h)$, to ensure the computed SOC for the time-step ahead SOC$(h+1)$ is feasible. The control filter takes into account the constraints on the battery storage's operation, such as maximum rate of charge or discharge ($\bar{p}^\mathrm{b+/-}$), battery's capacity and maximum depth of discharge. 



\begin{figure}[t]
\centering
\includegraphics[scale = 0.65]{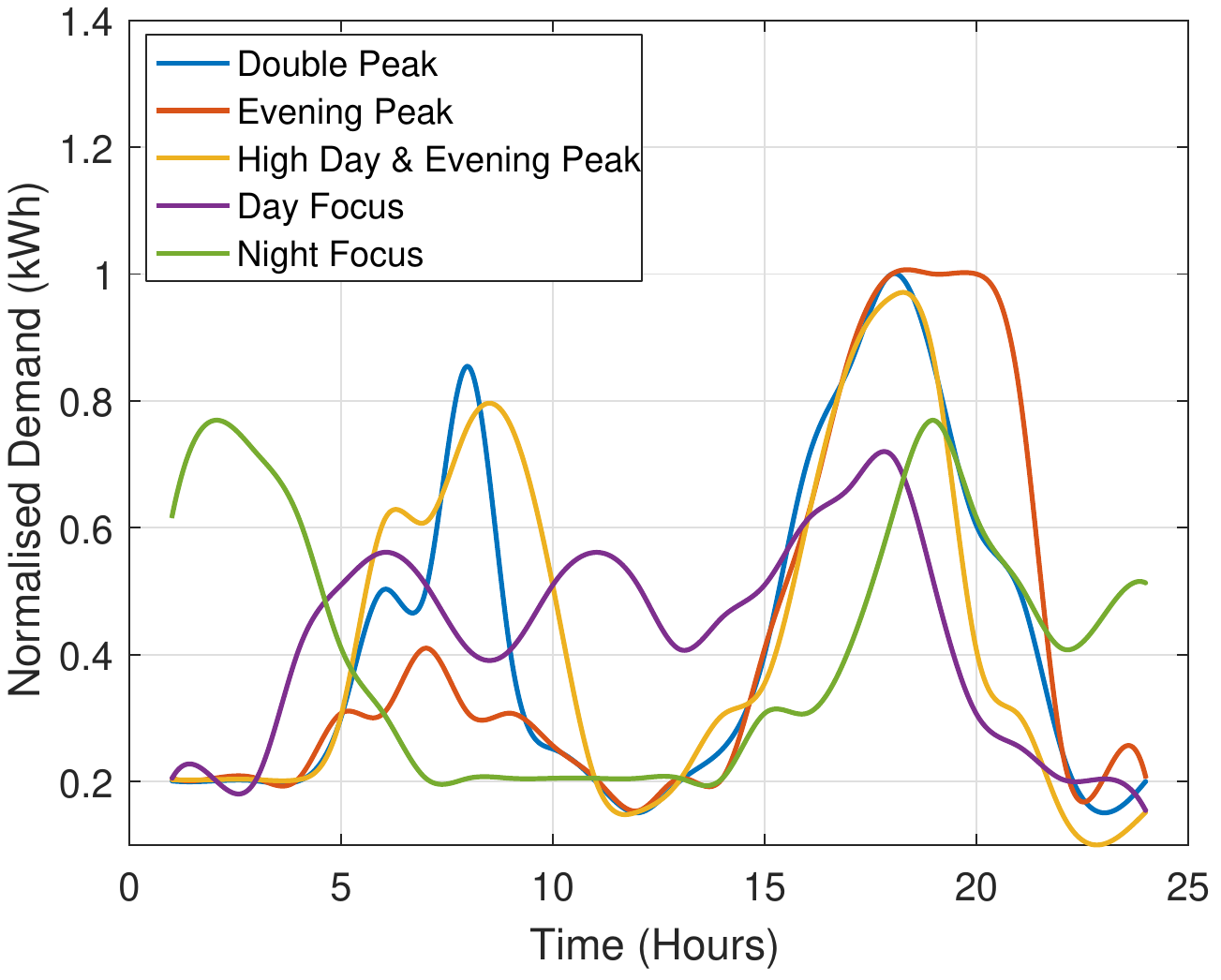}
\caption{Generic load profiles}
\label{fig6}
\end{figure}

\subsubsection{Policy function approximation using generic day types (PFAG)}
In this method, we cluster customer historic load data into five (5) generic load profile types described in~\cite{solarchoice}. The generic load profiles include \textit{Double Peak, Evening peak, High Day and Evening Peak, Day focus} and \textit{Night focus} and are shown in Figure~\ref{fig6}. They represent typical consumption patterns of residential customers in Australia depending on their load usage style and type or number of residents. Here, end-users are clustered, based on their half-hourly average consumption pattern during one day, for one year, and by applying a k-means clustering algorithm~\cite{2010:Na}. It is worth noting that the generic load profiles are considered as the centroids of each cluster.

As described in Algorithm~\ref{PlugPlayAlg}, for each cluster, a recurrent neural network (PFA) is trained for all end-users belonging to the particular cluster. The PFA is a time series nonlinear autoregressive (NARX) feedback neural network. In each cluster, the trained PFA for each end-user is tested on other end-users of the same cluster. The PFA with the least prediction error is selected as the representative PFA of the particular cluster. New end-users are then assigned to a cluster based on their consumption patterns, typically by filling-in a survey (Step 5). 
Then, the representative PFA is applied to a new user (Step 6) and a feasible SOC is computed by applying the control filter (Step 7)~\cite{paridari2018plug}.


\begin{algorithm}[t]
\caption{PFA using generic load type}
\label{PlugPlayAlg}
\begin{algorithmic}[1]
\State {Find end-users with considerable historical data.}
\State {Cluster them based on their consumption patterns.}
\State {Train PFA for end-users in each cluster.}
\State {Find the representative PFA for each cluster.}
\State {Assign new end-user into one of the clusters.}
\State {Apply representative PFA for new end-user.}
\State {Compute feasible SOC by applying control filter.}
\end{algorithmic}
\end{algorithm}

\subsection{Demand and PV persistence forecasting}	
We implement a simple persistence forecasting algorithm similar to that in~\cite{struth2013pv} in the following way for demand and PV prediction ($\tilde{P}_\mathrm{d}$  and $\tilde{P}_\mathrm{pv}$ respectively). This is done in order to investigate the performance of the energy management strategies with imperfect PV generation and demand forecast.

\subsubsection{Demand $\tilde{P}_{\emph{d}}$ prediction}
\noindent The weekday demand profiles of customers are based on their load profile a week before, to reflect the seasonality in their consumption pattern. Specifically, the predicted demand for the $i^\mathrm{th}$ time-step ahead is the same as the demand a week before at the same time step, i.e. $\tilde{P}_\mathrm{d}(h+i)=P_\mathrm{d}(h+i-7\ \mathrm{days})$.

\subsubsection{PV $\tilde{P}_{\emph{pv}}$ prediction}
\noindent Here, we set the predicted PV generation to be the same as the day before, but with a random variable $\xi$ to account for the prediction error. 
The random variable, assumed to follow a uniform distribution, is selected such that the predicted PV output deviates at most 10\% from the actual PV generation output. Therefore, the PV generation for the $i^\mathrm{th}$ time-step ahead is given as $\tilde{P}_\mathrm{pv}(h+i)=P_\mathrm{pv}(h+i)+\xi$.

\section{Economics} 
In this section we explain in details the parameters used in the financial calculations to measure the profitability of PV-battery systems. These include assumptions in electricity price and PV/battery market prices.

\subsection{Cost parameters}
The cost parameters used in the financial analysis are shown in Table~\ref{table4}. The initial investment cost of PV-battery systems are given as the total cost of the PV and battery systems as shown in Table~\ref{table5}~\cite{aginnovators}. We have assumed an annual electricity price increase of 3\% for the next 20 years~\cite{swoboda2014energy}. If this inflation is not catered for, there will be relatively lower internal rate of return (IRR) values.
%

\begin{table}[t]
	\footnotesize
	\centering
	\caption{Cost parameters}
	\label{table4}
	\begin{tabular}{lc}
		\hline \hline
		Cost Parameter & Value \\
		\hline
		Annual electricity price inflation, $e$ & 3\% \\
		Discount rate, $d$ & 5\% \\ 
		System lifespan, $\mathcal{N}$  & 20 years\\
		\hline
	\end{tabular}
\end{table}

\begin{table}[t]
	\footnotesize
	\centering
	\caption{PV-battery Market Prices}
	\label{table5}
	\begin{tabular}{l@{\hspace{0.2cm}}c@{\hspace{0.2cm}}c@{\hspace{0.2cm}}c@{\hspace{0.2cm}}c@{\hspace{0.2cm}}c@{\hspace{0.2cm}}c@{\hspace{0.2cm}}c@{\hspace{0.2cm}}c@{\hspace{0.2cm}}}
		\hline \hline
		\multicolumn{1}{c}{\multirow{2}{*}{\begin{tabular}[l]{@{}l@{}}Price\\ ($\times\ \$1000$)\end{tabular}}} & \multicolumn{8}{c}{PV-battery Sizes (kW/kWh)} \\ 
		\multicolumn{1}{c}{} & 3/6.5 & 4/6.5 & 5/9.8 & 6/9.8 & 7/14 & 8/14 & 9/14 & 10/14 \\ \hline
		PV & 4.4 & 5.3 & 6.1 & 7.5 & 8.9 & 10.3 & 11.7 & 13.1 \\ 
		Battery & 6.6 & 6.6 & 8.8 & 8.8 & 9.4 & 9.4 & 9.4 & 9.4 \\ 
		Total, $C_{0}$ & 11.0 & 11.9 & 14.9 & 16.3 & 18.3 & 19.7 & 21.1 & 22.5 \\ \hline
	\end{tabular}
\end{table}

\subsection{Financial indicators} \label{indices}
To assess the economic viability of PV-battery systems using the different energy management strategies, we employ two financial indicators namely, annual cost savings and internal rate of return (IRR).


\begin{itemize}
	\item Cost savings: To calculate the annual cost savings, we employ the formulas:
	\begin{equation} \label{eqn2}
    S_\mathrm{e} = C_\mathrm{e} - C_\mathrm{e}^\mathrm{DER}
	\end{equation}
	
	\begin{equation} \label{eqn3}
	C_{n} = S_\mathrm{e} + S_\mathrm{FiT}
	\end{equation}
	
	where:
	$S_\mathrm{e}$ is the annual electricity cost savings; 
	$C_\mathrm{e}$ is annual electricity cost without PV-battery;
	$C_\mathrm{e}^\mathrm{DER}$ is annual electricity cost with PV-battery (DER); 
	$C_{n}$ is total annual cost saving (cash inflow), and;
	$S_\mathrm{FiT}$ is the revenue from the FiT. \newline

	If inflation of electricity price is accounted for, the annual cash inflow escalates over the system lifespan. So, the equivalent discount rate $d'$, considering annual electricity price inflation $e$ is given by~\eqref{eqn4}~\cite{masters2013renewable}. Therefore, we find the levelized total annual cost savings by applying the \textit{levelising factor} (LF), given in~\eqref{eqn5}.
	
	\begin{equation} \label{eqn4}
	d'=  \frac{d-e}{1+e} 
	\end{equation}
		
	\begin{equation} \label{eqn5}
	LF=  \Bigg[\frac{(1 + d')^{n}-1}{d'(1 + d')^{n}} \Bigg] \cdot \Bigg[\frac{d(1 + d)^{n}}{(1 + d)^{n}-1} \Bigg]
	\end{equation}
	
	Hence, levelized total annual cost savings, $C'_{n}$ = $LF \cdot C_{n}$

	
	
	
	
	\item Internal rate of return (IRR): The internal rate of return $r$ is the discount rate at which NPV is zero. In other words, IRR measures how quick we break even or recover our initial investment cost. It is calculated by solving for $r$ in~\eqref{eqn8}. However, the IRR for the investment with electricity price inflation $r'$ is given by~\eqref{eqn9}:
	
	\begin{equation} \label{eqn8}
	\mathrm{NPV} =  - C_0 + \sum\limits_{n \in \mathcal{N}\setminus 0} \frac{C_{n}}{(1 + r)^{n}}= 0
	\end{equation}
    
	\begin{equation} \label{eqn9}
	r' =  r(1+e)+e 
	\end{equation}
	where: 
	$r$ is the internal rate of return (IRR) without inflation, and; 
    $C_0$ is the initial investment cost.
    
	
	
\end{itemize}

\tikzstyle{decision} = [diamond, draw, fill=blue!20, 
text width=4.5em, text badly centered, node distance=3cm, inner sep=0pt]
\tikzstyle{block1} = [rectangle, draw, text width=10em, text centered, rounded corners, very thick, minimum height=4em]
\tikzstyle{block2} = [rectangle, draw, text width=10em, text centered, rounded corners, very thick, minimum height=7em]
\tikzstyle{block3} = [rectangle, draw, text width=10em, text centered, rounded corners, very thick, minimum height=4em]
\tikzstyle{block4} = [rectangle, draw, text width=12em, text centered, rounded corners, very thick, minimum height=4em]
\tikzstyle{line} = [draw, -latex, thick]
\tikzstyle{cloud} = [draw, ellipse,fill=red!20, node distance=3cm,
minimum height=1em]
\begin{figure*}[t]
	\footnotesize
	\begin{center}
		\begin{tikzpicture}[node distance = 2cm, auto]
		\node[block1](input){\textbf{Idle aging influence \\ parameters} \\ SOC \\ Temperature};
		\node[block1, below of=input, node distance=2.2cm](input2){\textbf{Load aging influence \\ parameters} \\ Depth of cycling \\ C-rate \\ Ah-count \\ SOC limit \\ Temperature};
		\node[block2, below right = -0.8cm and 0.5cm  of input, node distance=2cm] (energy){\textbf{ Battery stress \\ characterization \\ method} \\ Half-cycle counting \\ (event-based counting)};
		\node [block3, below right = -1.95cm and 0.5cm of energy, node distance=3.5cm] (life) {\textbf{Battery lifetime \\ models} \\ Calendrical aging \\ $\boldsymbol{+}$ \\ Cyclical aging };
		\node [block4, below right = -1.83cm and 0.5cm of life, node distance=3.5cm] (capacity) {\textbf{Capacity and  performance \\ degradation} \\ SOH \\ Expected lifetime\\Annual full equivalent cycles};

		\draw[line, very thick] (1.5,-0.18) node[right]{}  -- (2,-0.18);
		\draw[line, very thick] (1.5,-1.7) node[right]{}  -- (2,-1.7);
		\draw[line, very thick] (5.07,-0.8) node[right]{}  -- (5.55,-0.8);
		\draw[line, very thick] (8.6,-0.85) node[right]{}  -- (9.1,-0.85);
		\end{tikzpicture}
		\caption{SimSES battery degradation model~\cite{naumann2017simses}.}
	    \label{figure7}
	\end{center}
\end{figure*}
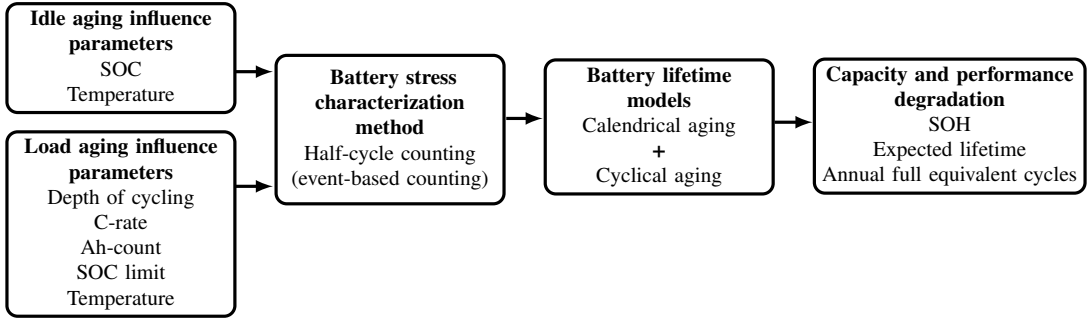

\section{Battery degradation study}
To assess the battery degradation over the system lifetime, we use the SimSES (software for techno-economic simulation of stationary energy storage systems) open-source software~\cite{naumann2017simses}. It enables a detailed simulation and evaluation of stationary energy storage systems, particularly lithium-ion batteries. The SimSES battery degradation model, depicted in Figure~\ref{figure7}, receives the battery \textit{Idle} and \textit{Load} aging input parameters and implements the half-cycle counting battery stress characterization method to estimate the battery calendrical and cyclical aging. Next, the combined effects of the calendar (idle stress) and cyclic (load stress) aging are used to estimate the battery capacity degradation and expected lifetime~\eqref{eqn15}. The yearly battery charge/discharge power and SOC, which are outputs of the energy management simulation, along with other aging parameters given in~\cite{naumann2017simses}, are the inputs to the model. The aging influence parameters include the battery depth of cycling (DOC), C-rate, Ah-count, SOC limit, and temperature while the output of the model includes the battery state of health (SOH) after 20 years, the average annual full equivalent cycles and the expected lifetime (at 80\% SOH). 

Formally, following~\cite{murnane2017closer,truong2016economics}, the SOH is estimated by:     
\begin{equation} \label{eqn11}
\mathrm{SOH} = \frac{C_\mathrm{max}}{C_\mathrm{rated}}\cdot 100\%
\end{equation}
where:
$C_\mathrm{max}$ is the maximum releasable battery capacity (which declines with time), and; 
$C_\mathrm{rated}$ is the battery rated capacity.
Calendric degradation is given by:
\begin{equation} \label{eqn12}
\Delta C_\mathrm{cal} =  \frac{0.2 \cdot C_\mathrm{rated}}{t_\mathrm{cal}}
\end{equation}
where: 
$\Delta C_\mathrm{cal}$ is capacity degradation due to calendric aging, and; 
$t_\mathrm{cal}$ is calendric time period until battery degrades by 20\% of its rated capacity.
Cyclic capacity degradation is given by:
\begin{equation} \label{eqn13}
\Delta C_\mathrm{cyc} =  \frac{0.2 \cdot C_\mathrm{rated}}{k_\mathrm{cyc}(DOC)\cdot DOC}
\end{equation}
where: 
$\Delta C_\mathrm{cyc}$ is capacity degradation due to cyclic aging, and; 
$k_\mathrm{cyc}$ is amount of equivalent full cycles until battery degrades by 20\% of its rated capacity.

Given the values in~\eqref{eqn11},~\eqref{eqn12} and~\eqref{eqn13}, the battery aging model of ~\cite{benjaminb} is given by:
\begin{equation} \label{eqn14}
	a(t) = a_0 + \sum\limits_{t \in \mathcal{T}}da(t)
\end{equation}
\begin{equation} \label{eqn15}
	da(t) = da_\mathrm{cal}(t) + da_\mathrm{cyc}^{+}(t) + da_\mathrm{cyc}^{-}(t)
\end{equation}
\begin{equation} \label{eqn16}
v(t) = 1 - (1 - v_\mathrm{e})\ a(t)
\end{equation}
where: 
$a(t)$ is the battery age;  
$a_0$ is the aging scaling constant;  
$da_\mathrm{cal}(t)$ is the battery calendric aging;  $da_\mathrm{cyc}^{+/-}(t)$ is the battery cyclic (charge/discharge) aging; 
$v(t)$ is the normalised battery capacity; 
$v_\mathrm{e}$ is the normalised battery capacity at end of life, and; $\mathcal{T}$ is the total simulation time.

\begin{figure*}[!htb] 
\centering
	\hspace{-0.8em}
	\subfloat[MILP]{%
		\includegraphics[scale = 0.5]{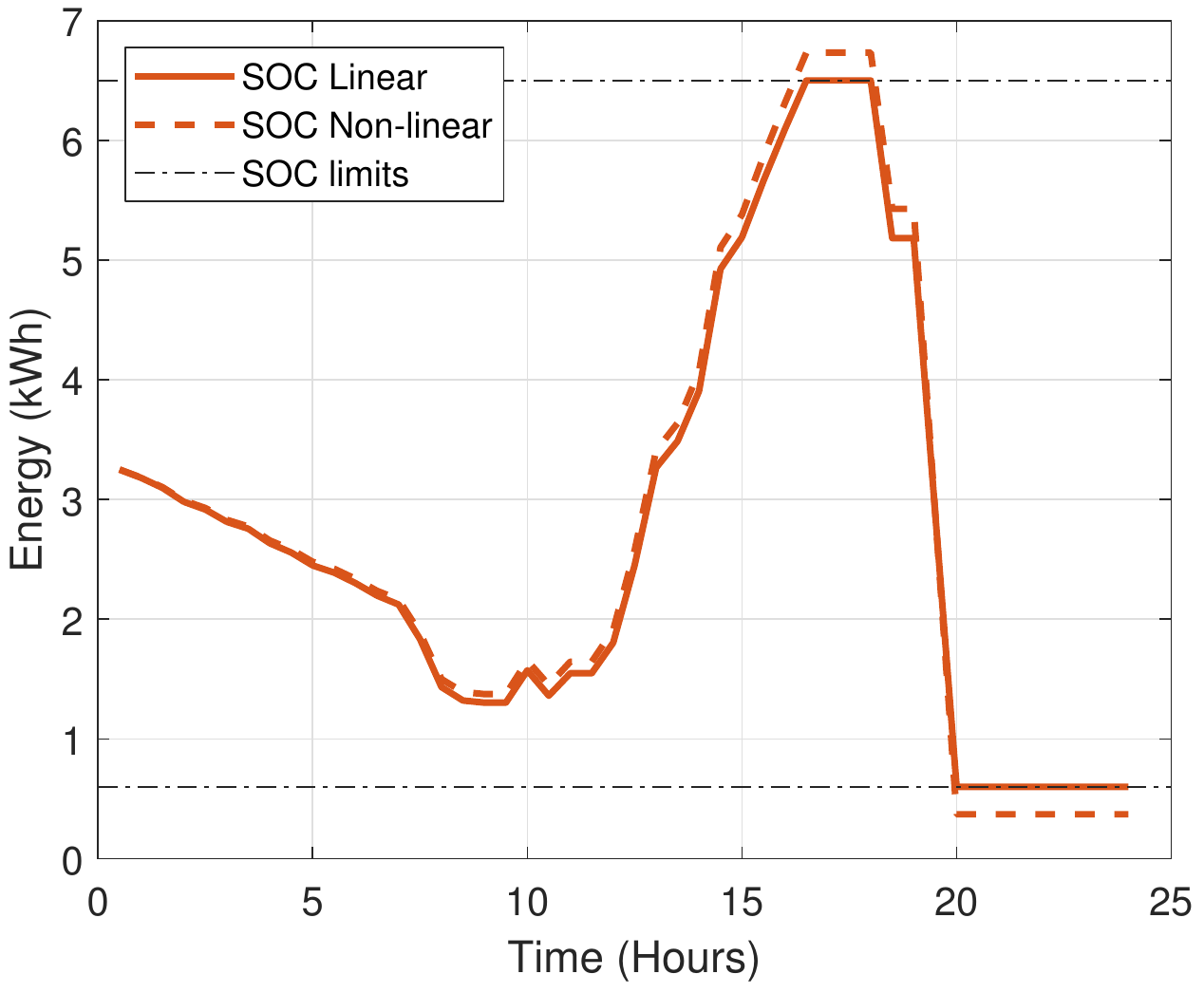} \label{figure8a}
	}  \hspace{1.3em}
	\subfloat[SCM]{%
		\includegraphics[scale = 0.5]{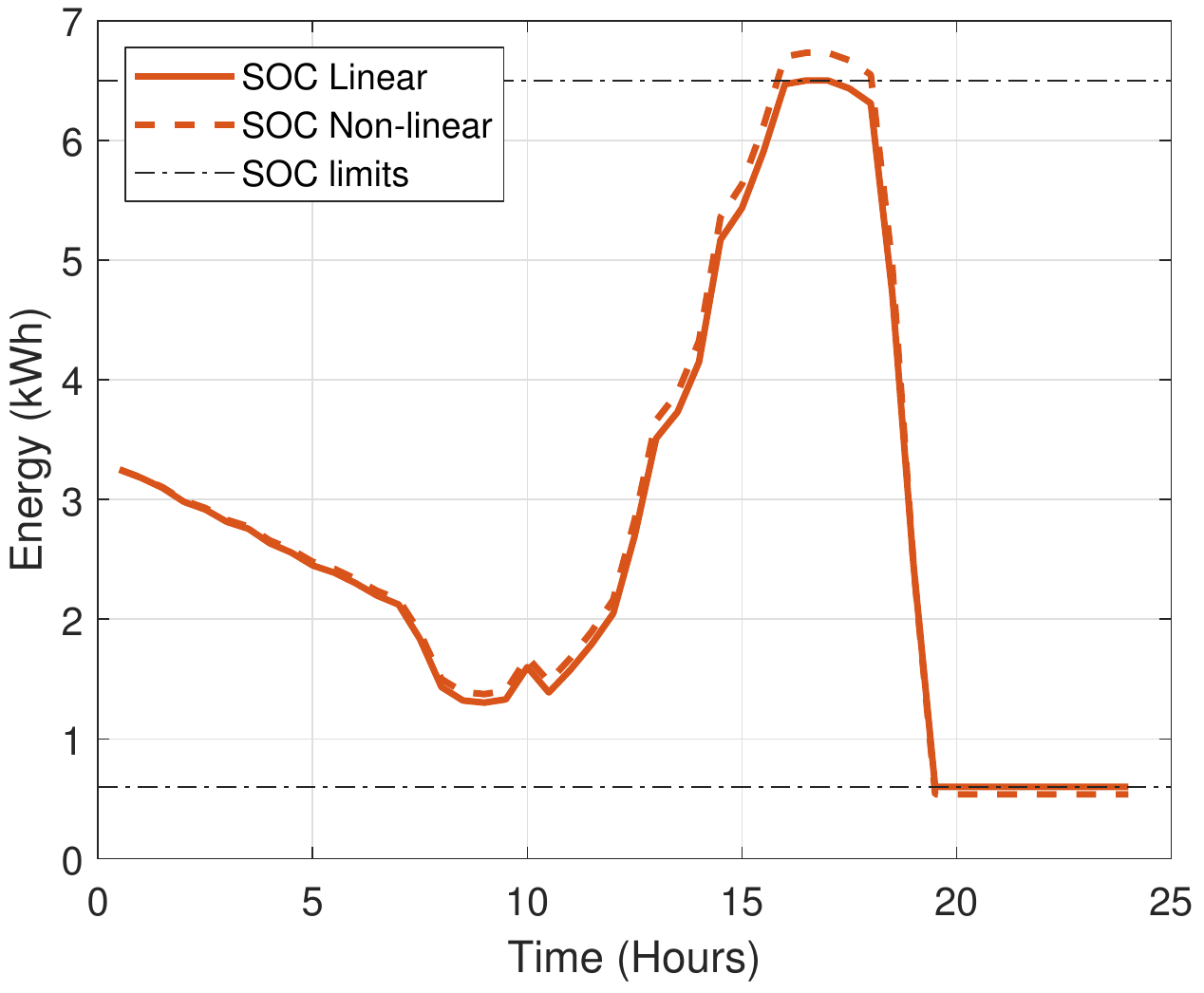} \label{figure8b}
	} 
	\caption{SOC plot with linear and nonlinear battery models for MILP and SCM energy management strategies (Customer 26, Day 1).}
	\label{figure8}
\end{figure*}

\section{Results and analysis}
In this section, we present the simulation results of the energy management strategies in the following order. First, we show the energy scheduling plots, followed by the economic assessment results. Third, we discuss the computational requirements of each method and finally, we present the results from the battery degradation study.

\subsection{Scheduling results} \label{scheduling}
Here, we show the energy scheduling (battery scheduling  and grid power exchange) results of the different energy management strategies. In Figure~\ref{figure8}, we show the difference in battery SOC that results when the MILP or SCM battery charging power profile (obtained using a linear battery model) is applied to a nonlinear battery model instead of a linear one. This highlights the fact that MILP and SCM results can be infeasible (SOC limit constraint violated), which in practice eventually leads to an overestimation of the HEMS economic performance, as discussed in Section~\ref{batterymodel}. 

For illustration purposes, we show in Figure~\ref{figure9} how the household demand is met from different power sources (PV, battery or grid), and how the PV power is utilised, with the MILP energy management strategy while in Figure~\ref{figure10}, we show the scheduling results of a randomly selected customer (Customer 47) on the third day of the year. The battery scheduling with MILP (Figure~\ref{figure10a}) and SCM (Figure~\ref{figure10f}) are similar, since MILP also maximises self-consumption due to the low FiT rate relative to the electricity retail price. For the period between 07:00 -- 08:00 and 17:00 -- 22:00, the household demand is met using the battery power as depicted in Figure~\ref{figure9}. 

\par However, there is a difference between SCM and MILP scheduling. With SCM, grid power import is done only when PV and battery power is unavailable while with MILP, the decision to import or export power is done relative to cost minimisation. In ToUA (Figure~\ref{figure10e}), the algorithm ensures a certain battery power (in this case, 30\% of the maximum battery SOC) is left at the end of the day, by charging the battery with cheap off-peak grid power. And at the beginning of the day, the battery will be idle if discharging will cause its SOC to fall below the 30\% SOC threshold, until the start of the high price (off-peak and shoulder) electricity periods. This is a form of energy security, which is useful when there is a day-ahead forecast of low PV generation. 

\par Similar to Figure~\ref{figure10a} (MILP), Figure~\ref{figure10b} (energy scheduling plot using DP) is an outcome of a cost minimisation with respect to time differentiated prices. Here, the battery charges with cheap off-peak power and discharges when the electricity prices are high. However, unlike the MILP and other energy management strategies, the DP approach uses a more accurate battery operating model. This model better represents reality than the other strategies since it captures the nonlinearities in the battery and inverter operation, hence, the distinct difference in the battery scheduling results. 

\par With PFAS and PFAG, battery scheduling is the output of a neural network trained using the MILP scheduling results. Since PFAS is tailored specifically to individual customers, its energy scheduling plot (Figure~\ref{figure10c}) is a smoothed version of that of MILP. With PFAG (Figure~\ref{figure10d}), however, a neural network representative of a cluster is used to predict the energy scheduling output of customers belonging to the particular cluster. In this case, Cluster 4 representative PFA is used to predict the results of Customer 47. Hence, the battery scheduling with PFAG and PFAS are substantially different, even for the same PV and demand input data.

\begin{figure}[t]
	\centering
	\includegraphics[scale = 0.55]{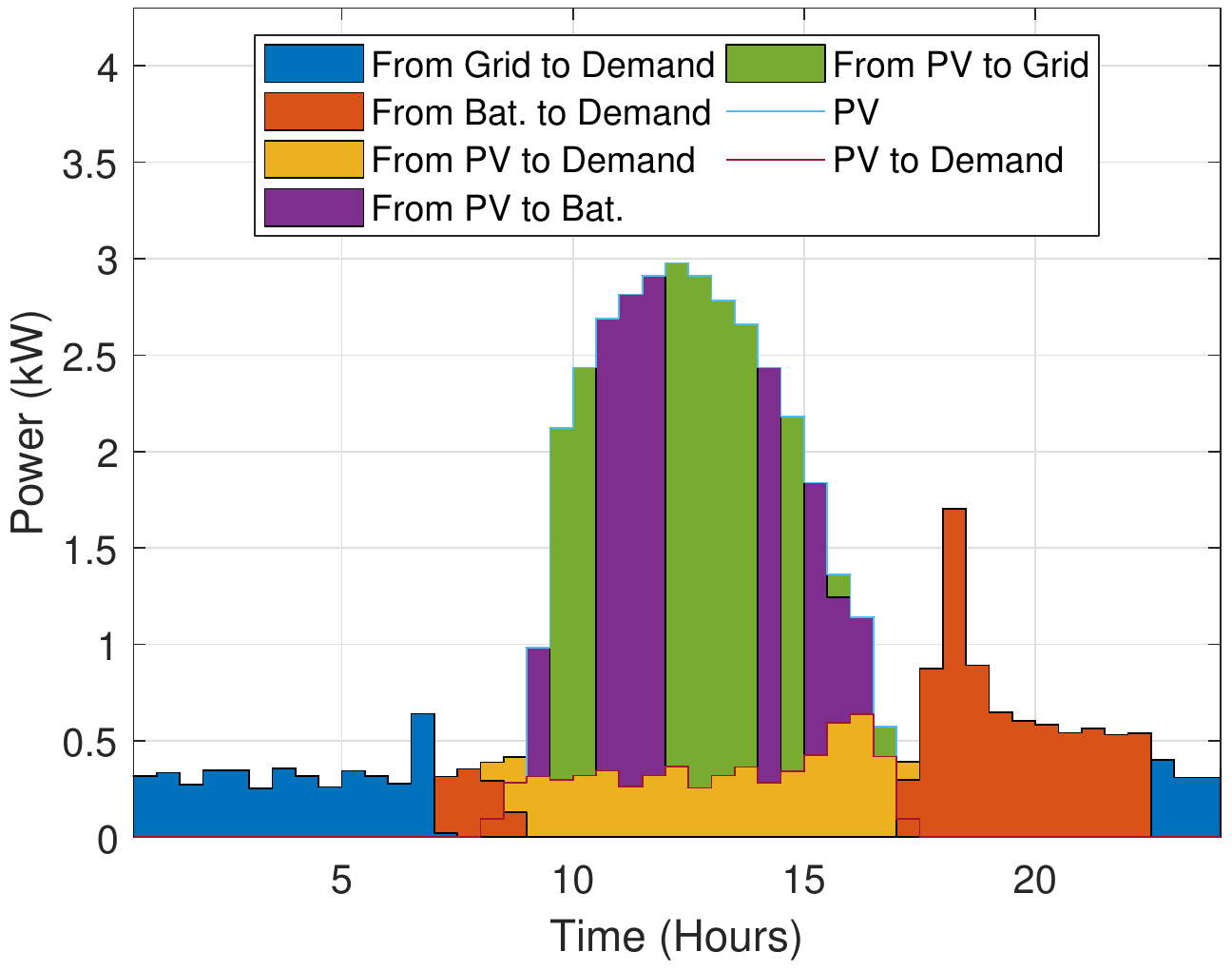}
	\caption{Household power flows with MILP for Customer 47, Day 3.}
	\label{figure9}
\end{figure}


\begin{figure*}[!htb] 
\centering
	\hspace{-0.8em}
	\subfloat[MILP]{%
		\includegraphics[scale = 0.55]{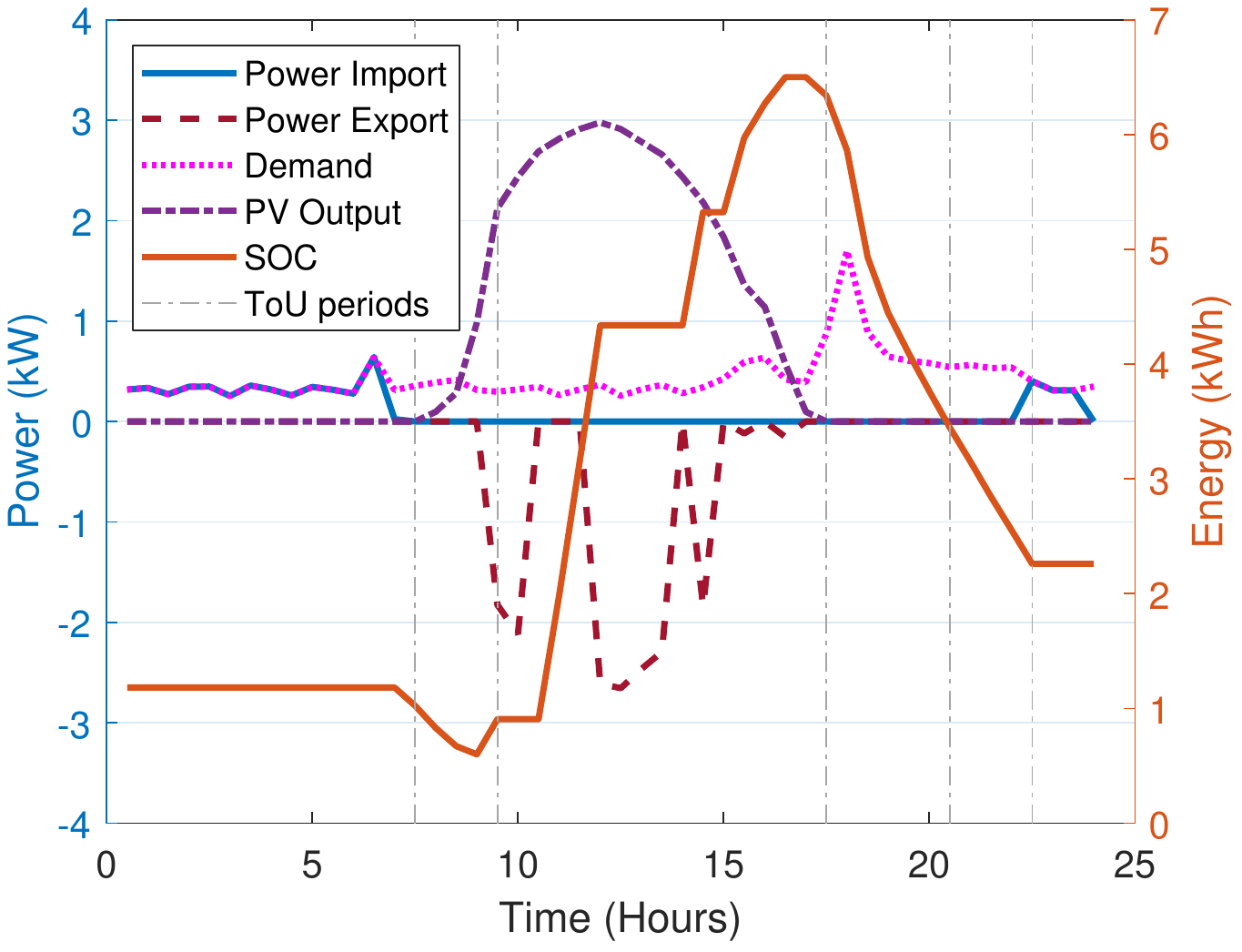} \label{figure10a}
	}  \hspace{1.3em}
	\subfloat[DP]{%
		\includegraphics[scale = 0.55]{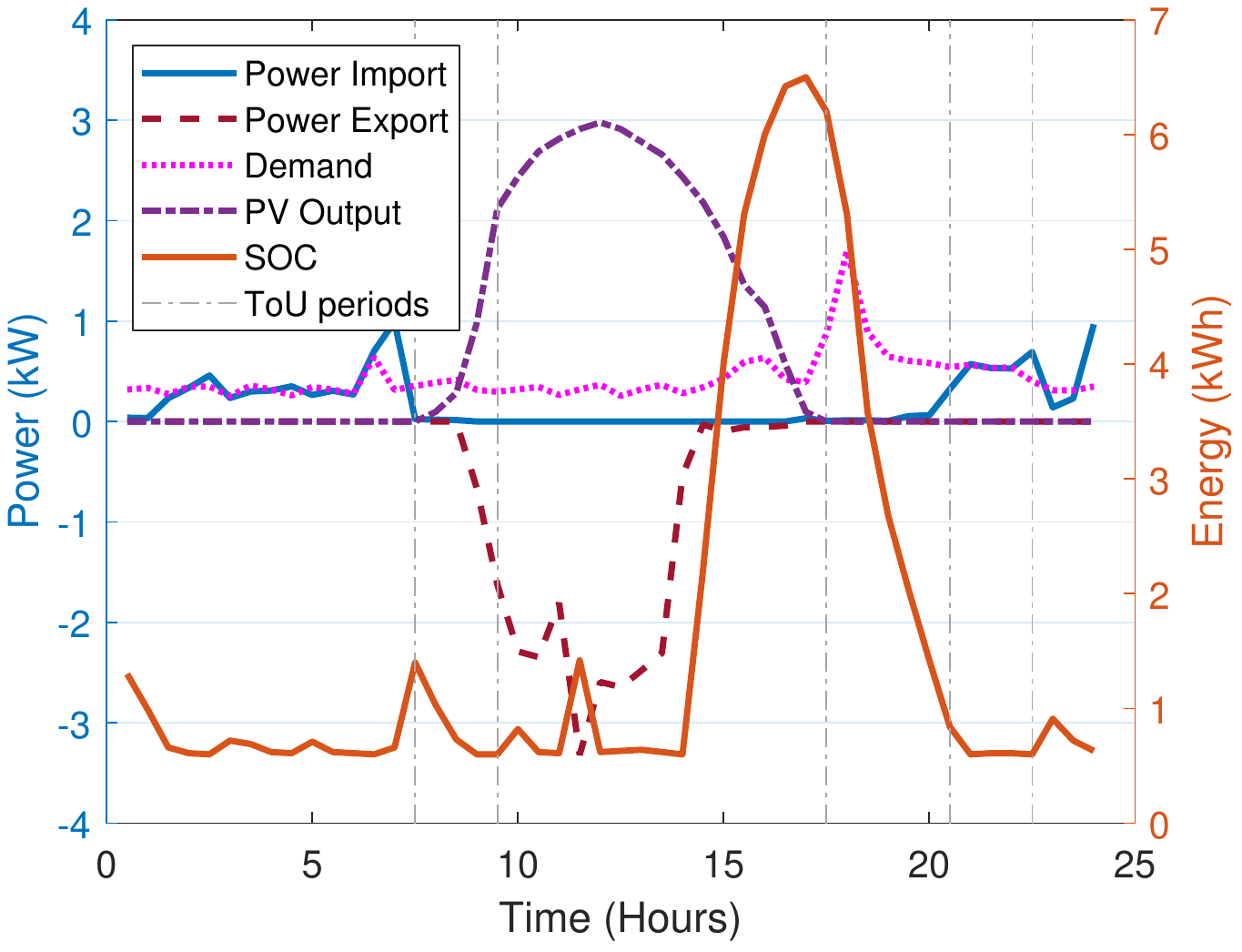} \label{figure10b}
	} 
	\vspace{-0.08em}
	\subfloat[PFAS]{%
		\includegraphics[scale = 0.55]{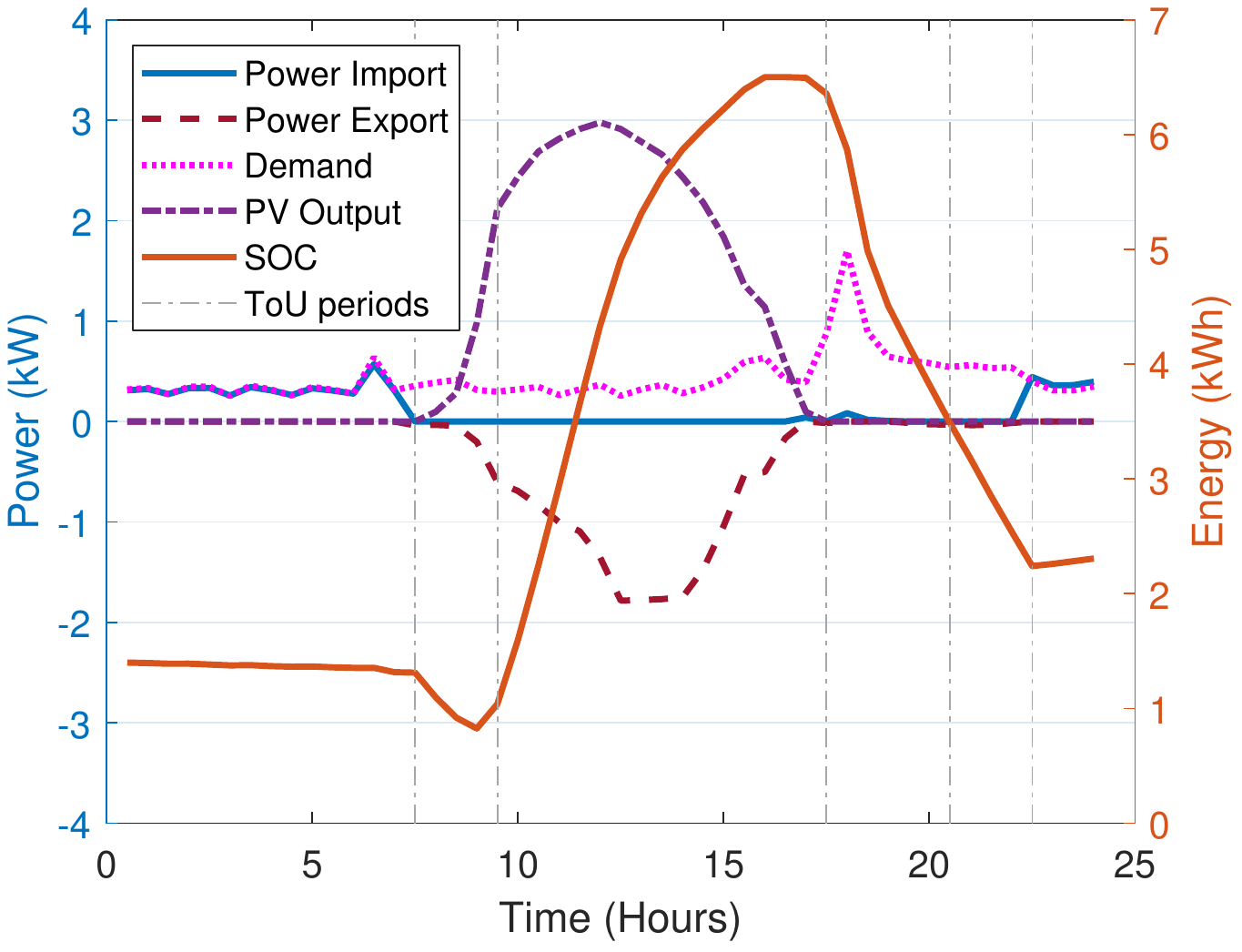} \label{figure10c}
	}  \hspace{1.3em}
	\subfloat[PFAG]{%
		\includegraphics[scale = 0.55]{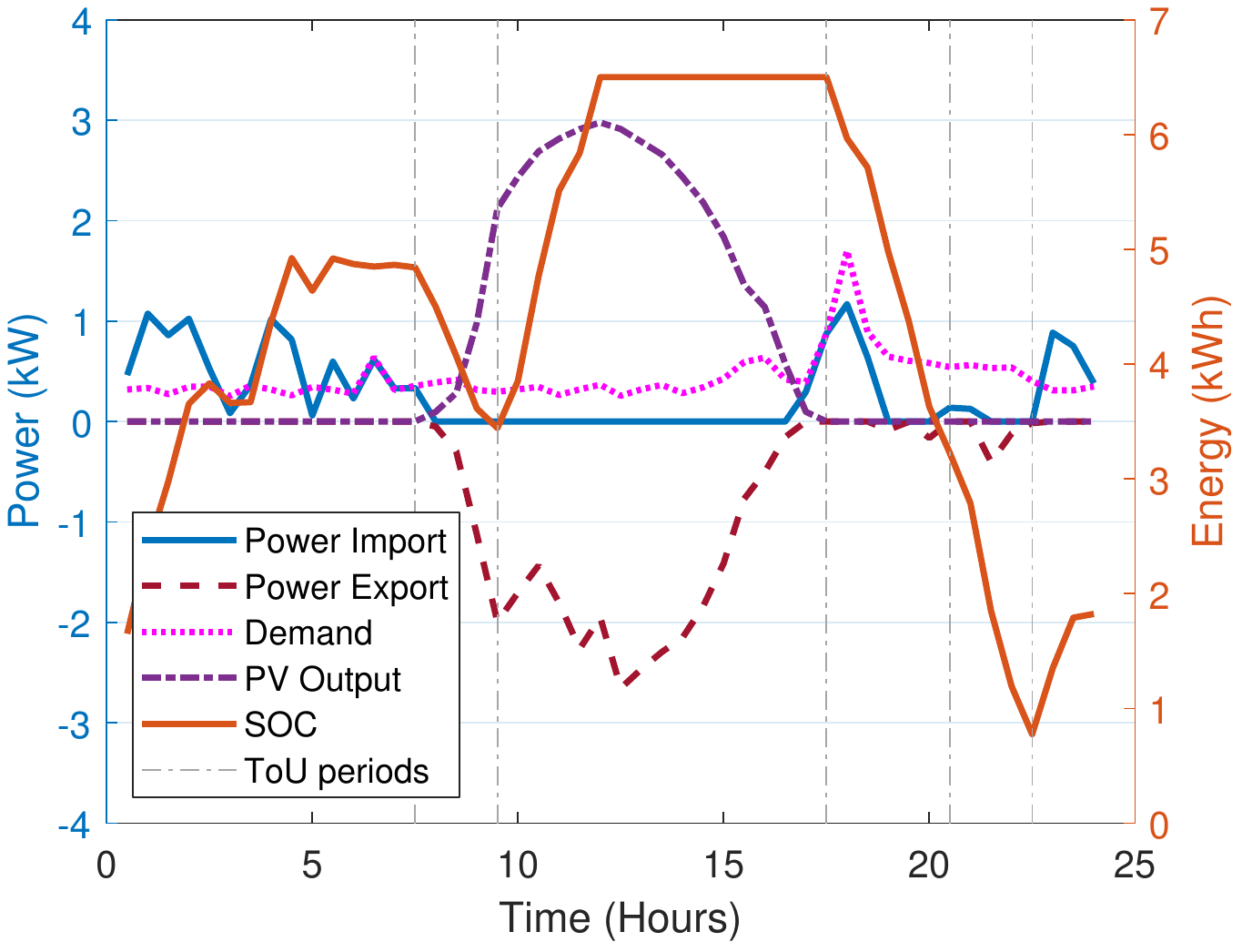} \label{figure10d}
	} 
	\vspace{-0.08em}
	\subfloat[ToUA]{%
		\includegraphics[scale = 0.55]{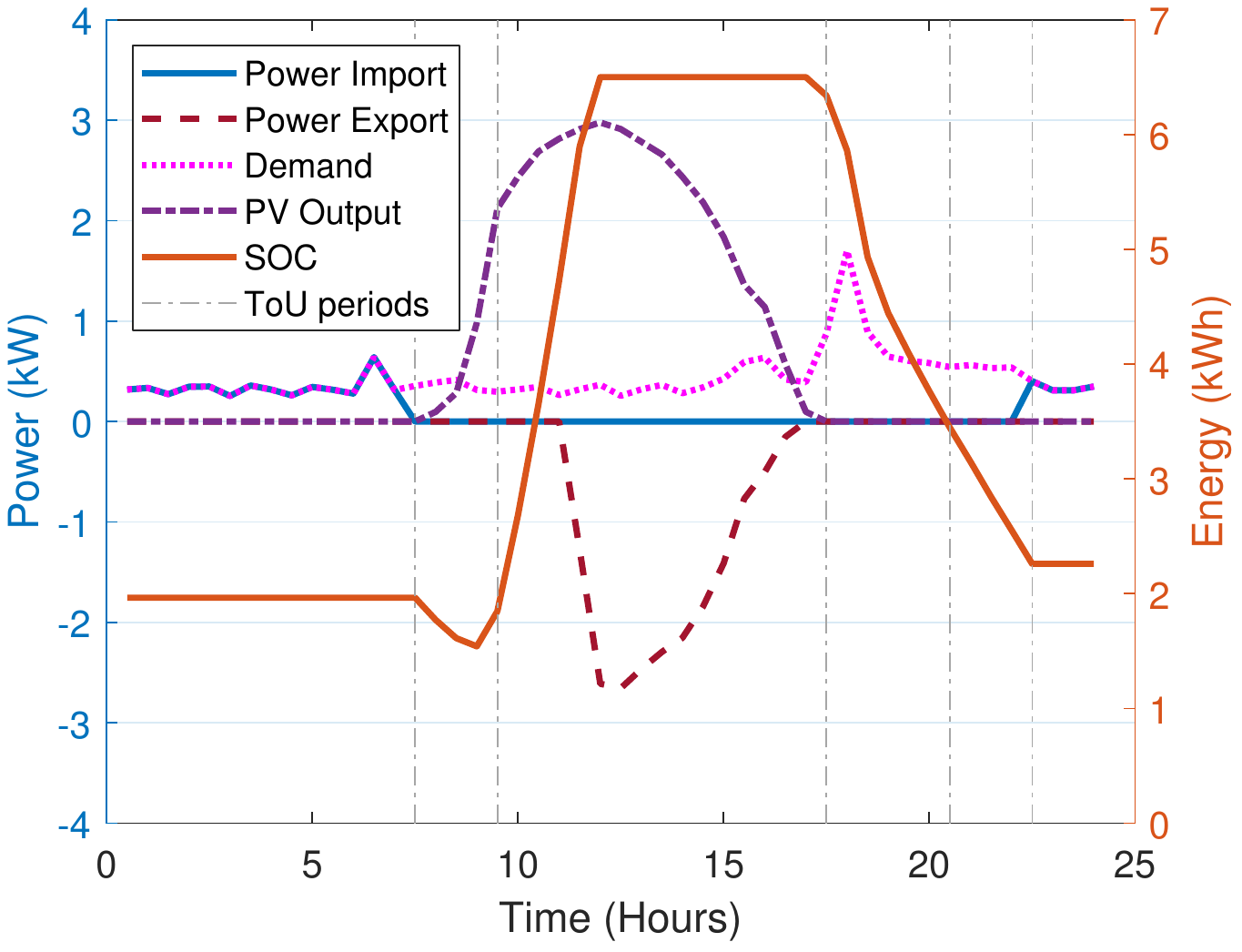} \label{figure10e}
	} \hspace{1.38em}
	\subfloat[SCM]{%
		\includegraphics[scale = 0.55]{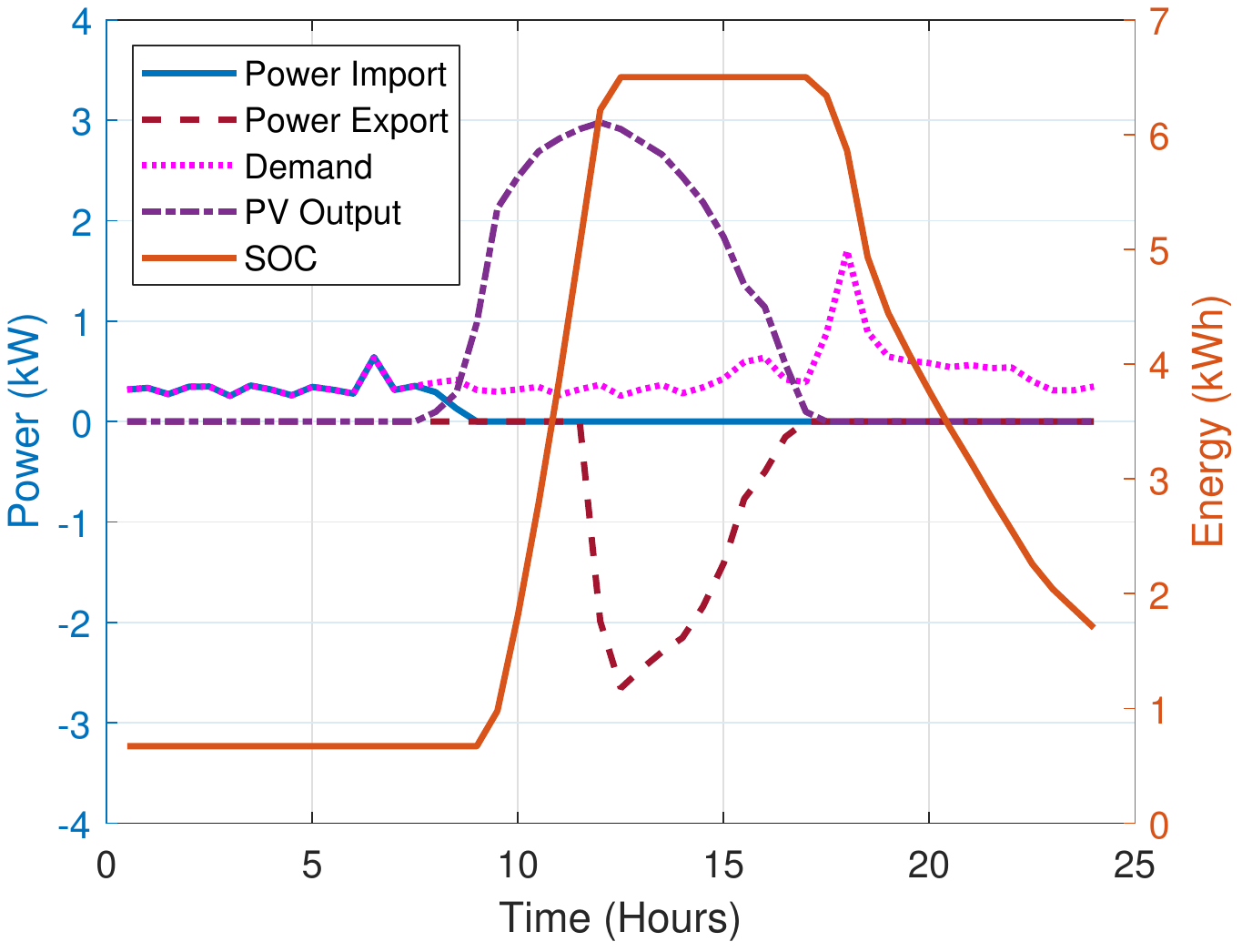} \label{figure10f}
	}
	\caption{Daily power flows and battery schedules for Customer 47, Day 3.}
	\label{figure10}
\end{figure*}

\begin{figure*}[!htb] 
\centering
	\hspace{-0.8em}
	\subfloat[Annual cost savings]{%
		\includegraphics[scale = 0.65]{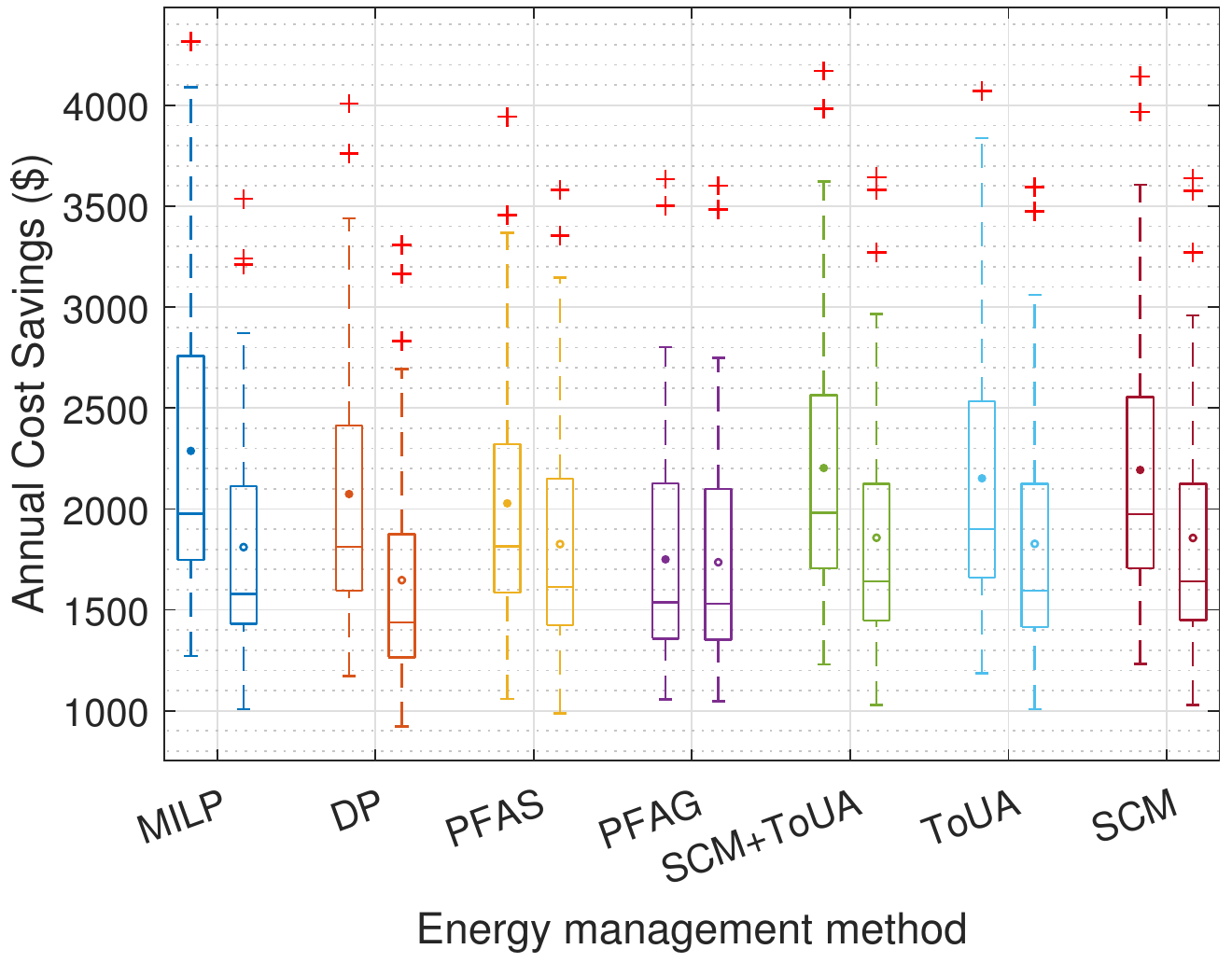} \label{figure11a}
	}  \hspace{1.3em}
	\subfloat[Internal rate of return (IRR)]{%
		\includegraphics[scale = 0.65]{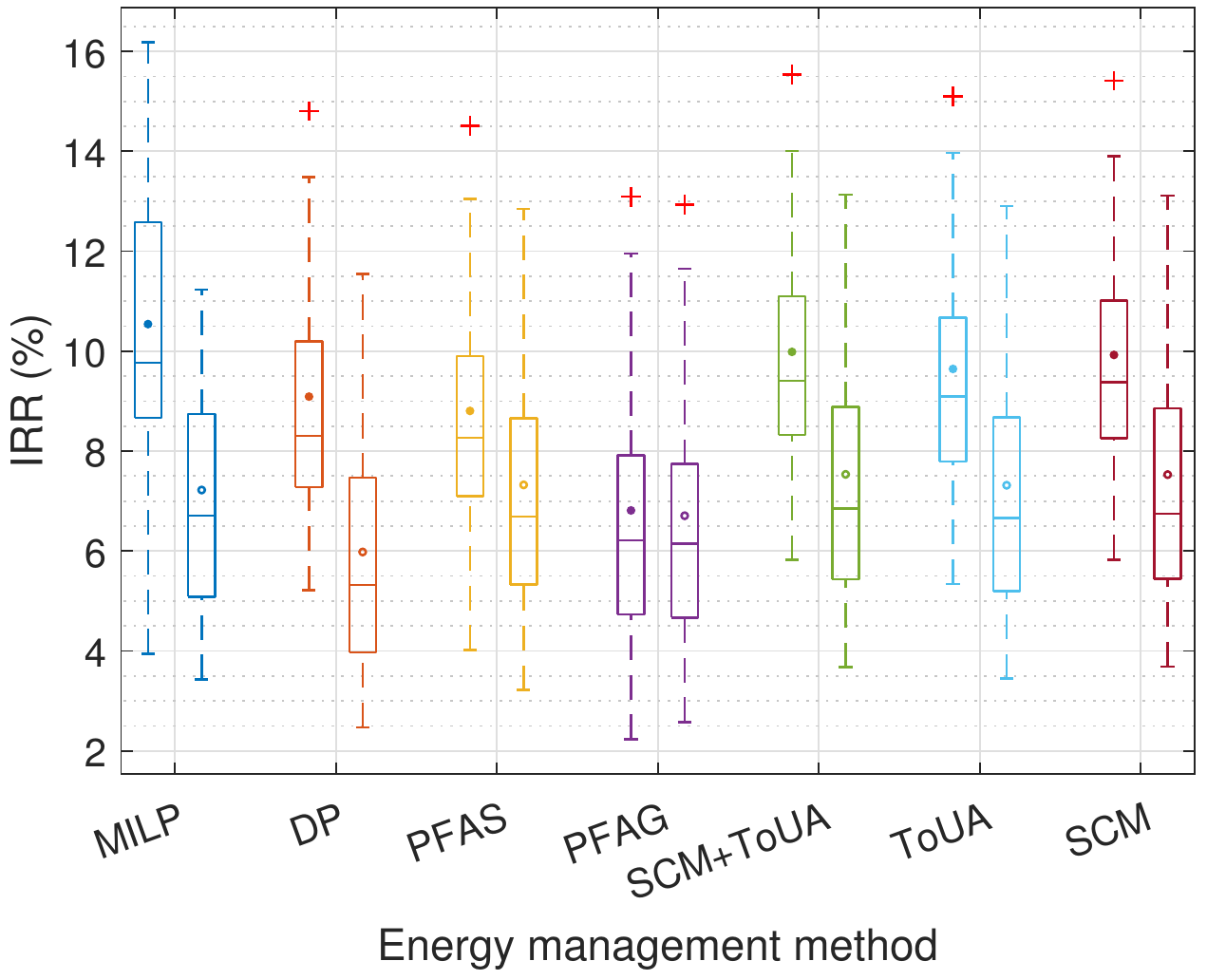} \label{figure11b}
	} 
	\caption{Financial indices for all 52 customers considering perfect ($\color{black}{\bullet}$) and imperfect forecast ($\color{black}{\boldsymbol{\circ}}$). `$\color{red}{+}$' indicates box plot outliers.}
	\label{figure11}
\end{figure*}

\subsection{Economic viability} \label{ecoviab}
Using the financial indicators described in Section~\ref{indices}, we now perform an economic assessment of the PV-battery system under the different energy management strategies in this section. The plots in Figure~\ref{figure11} show the statistical results of the financial indicators for all 52 customers. It is worth noting that the results from MILP and the rule-based heuristics are based on a linearised battery model, and therefore can be overly optimistic.

\par To begin with, we compare MILP with the rule-based heuristic strategies. In Figure~\ref{figure11a}, the total annual cost savings (median value) assuming perfect forecast of PV and demand is the highest with MILP, followed by SCM+ToUA and SCM, both of which are relatively close. Amongst the heuristic strategies, ToUA has the least cost savings and a similar pattern follows for IRR. This is expected, because performing arbitrage on a daily basis is not economically worthwhile if the stored battery power is not effectively utilised due to available PV generation during the day. 


\par However, with imperfect forecast, the performance of MILP deteriorates.  Generally, because MILP is a principled optimisation method that minimises cost, it's performance will be more adversely affected with imperfect forecast when compared with the heuristic strategies, especially with imperfect price response using ToU tariffs. SCM and SCM+ToUA both perform similarly and results in more cost savings compared to MILP and ToUA. This shows that with imperfect forecast (and no additional controllers on the battery operation under MILP), the best strategy tends to be to maximise PV self-consumption.

\par Next, we discuss the performance of MILP relative to PFAS, PFAG and DP. Although, the machine learning approaches (PFAS and PFAG) were developed based on the MILP optimisation results, they show more robustness to imperfect PV and demand forecasts, compared to the principled optimisation approaches. This is due to the cost minimisation strategy embedded in these approaches. Since PFAS is customer specific, it is expected to outperform PFAG. However, the advantage of PFAG over PFAS, is that it requires fewer prior customer information/data. It is essentially a plug-and-play strategy. On the other hand, while MILP appears to perform better than DP with perfect and imperfect forecasts in simulations, this is not the case in practice. This is because MILP uses a linearised battery and inverter model and as such does not capture the nonlinearities unlike in DP, where the nonlinear transitions are accurately modelled. Therefore, on implementation, DP will outperform MILP and all the other strategies where a linear battery model is assumed.

\par A summary of the financial performance of each strategy is given in Table~\ref{table7} in form of rankings. Since these are only simulation results, we will unpack further in the conclusions, the economic performance of each strategy in the light of practical implementation. 


\subsection{Computation time}  \label{comptime}
\par With reference to Tables~\ref{table6} and~\ref{table7}, we can conclude that although the MILP algorithm results in the most cost savings (considering a perfect forecast of PV and demand), the heuristic strategies will be preferred to the MILP algorithm in terms of the computational performance. The SCM algorithm solves the daily energy management problem for a customer in less than 0.006 seconds while ToU takes twice as much to execute same task. Although MILP takes 0.084 seconds (15 times slower than SCM) to solve same problem, it is not a barrier to implementing MILP since other principled optimisation approaches like DP take much longer~\cite{keerthisinghe2016fast} (about 30 seconds in our case study). This is because a linearised model of the battery operation has been assumed for the rule-based heuristic strategies and MILP. This makes for simpler models with lower computational burden, compared to the DP (see Table~\ref{table7}). 

\par Similarly, the machine learning approaches take a reasonable amount of time to solve the same problem, while training is only done once. PFAG takes approximately 300 seconds for training and nearly 0.6 seconds to execute the energy management for a customer in a day. However, in order to improve the performance of PFAG, we run the RNN prediction for 20 time steps ahead in closed loop mode, thereby creating an additional neural network that takes an extra 7 seconds to execute. PFAS on the other hand, takes less than half the time for training (92 seconds) and executing (0.0118 seconds) the same task. Likewise, in order to improve its performance, we used a SOC feedback loop to carry out time series prediction and policy filtering at each time step. For this additional step, an extra 52 seconds is required for training, while execution takes 0.26 seconds. Furthermore, PFAS gives a faster and more accurate solution compared to PFAG since it is customer specific. Since we have carried out all our simulations in MATLAB, it is worth mentioning that the PFA strategies will be much faster than optimisation approaches like MILP when implemented in programming languages like Python due to its superior neural network libraries.

Generally, to make a fair comparison with regards to the computational performance, we need to consider the operation of these energy management strategies with a nonlinear battery model. Apart from the DP method, all other strategies considered in this work assume a linear battery and inverter operating model. In this case, the computational speed of MILP and the rule-based heuristics will be adversely affected. For example, the MILP strategy would have to be run several times to account for the nonlinearities, thereby increasing its computation time. However, the machine learning approaches can be trained using the DP results, therefore inherently capturing the nonlinearities in the battery and inverter operation without a deterioration in their computational performance. Therefore, the results in Table~\ref{table7} for computational speed holds true on the assumption of a linear battery model.



\begin{table}[t]
	\footnotesize
	\centering
	\caption{Daily computation time for one customer}
	\label{table6}
\begin{tabular}{ccc}
\hline \hline
\multirow{2}{*}{\begin{tabular}[c]{@{}c@{}}Energy \\ management\\  strategy\end{tabular}} & \multicolumn{2}{c}{Computation time (s)} \\ 
 & \begin{tabular}[c]{@{}c@{}}Training\\ (offline)\end{tabular} & \begin{tabular}[c]{@{}c@{}}Execution \\ (online)\end{tabular} \\ \hline
SCM & - & 0.005548 \\ 
ToUA & - & 0.014576 \\ 
SCM+ToUA & - & 0.008863\\ 
PFAG (RNN) & 300  & 0.585706  \\ 
PFAS (ANN) &  92 & 0.011801  \\ 
DP & - & 30.199325  \\
MILP & - & 0.084133  \\  \hline
\end{tabular}
\end{table}


\begin{figure*}[!htb] 
\centering
	\subfloat[Annual full equivalent cycles (FEC)]{%
		\includegraphics[scale = 0.65]{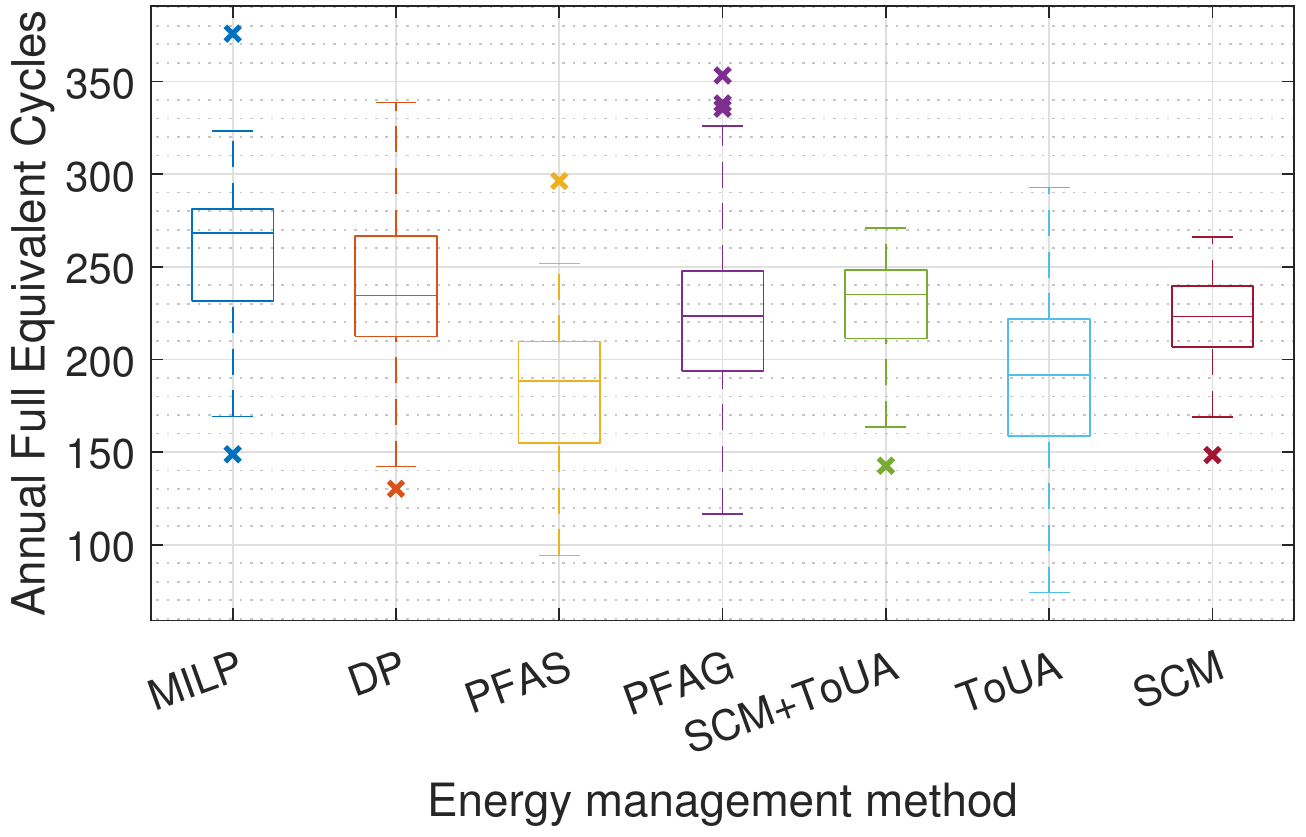} \label{figure12a}
	}  \hspace{1.4em}
	\subfloat[State of health after 20 years (SOH)]{%
		\includegraphics[scale = 0.65]{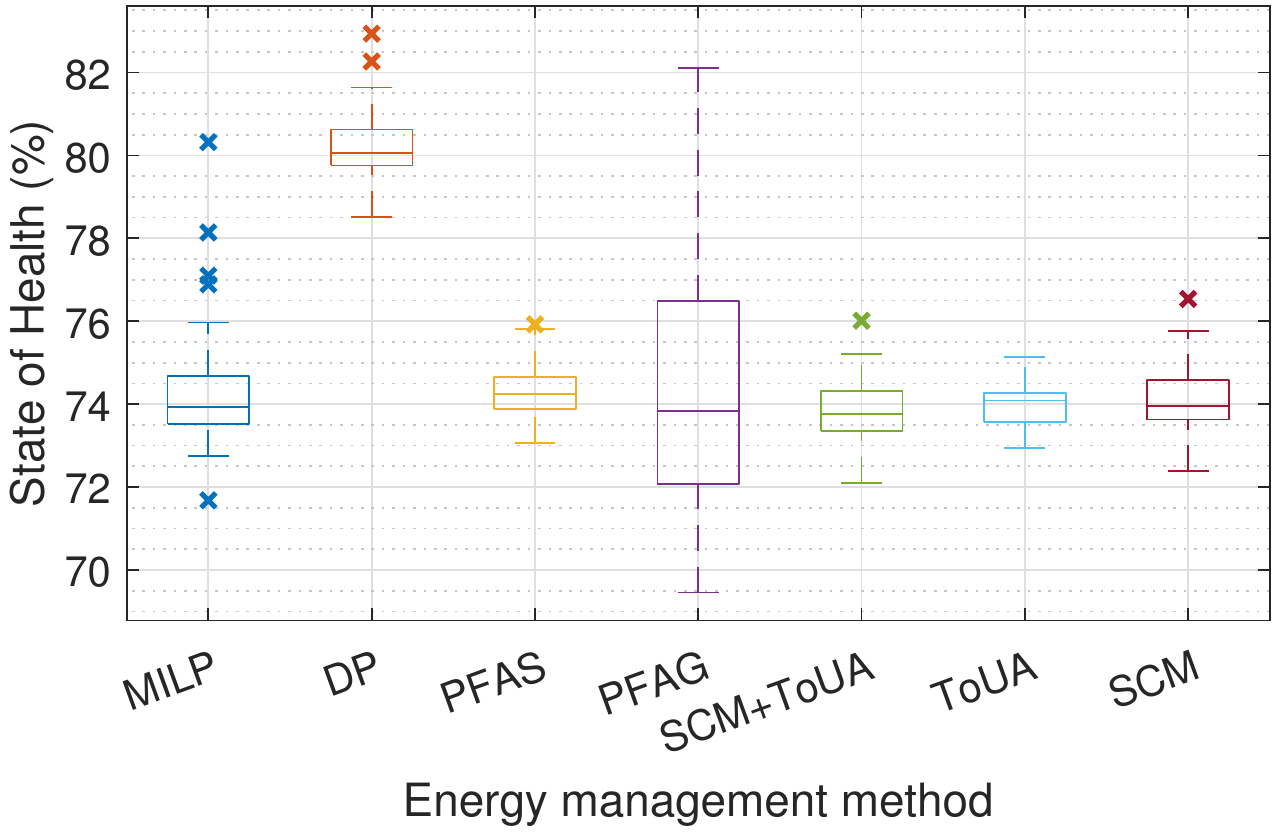} \label{figure12b}
	} 
	
	\subfloat[Expected battery lifetime in years (EBL)]{%
		\includegraphics[scale = 0.65]{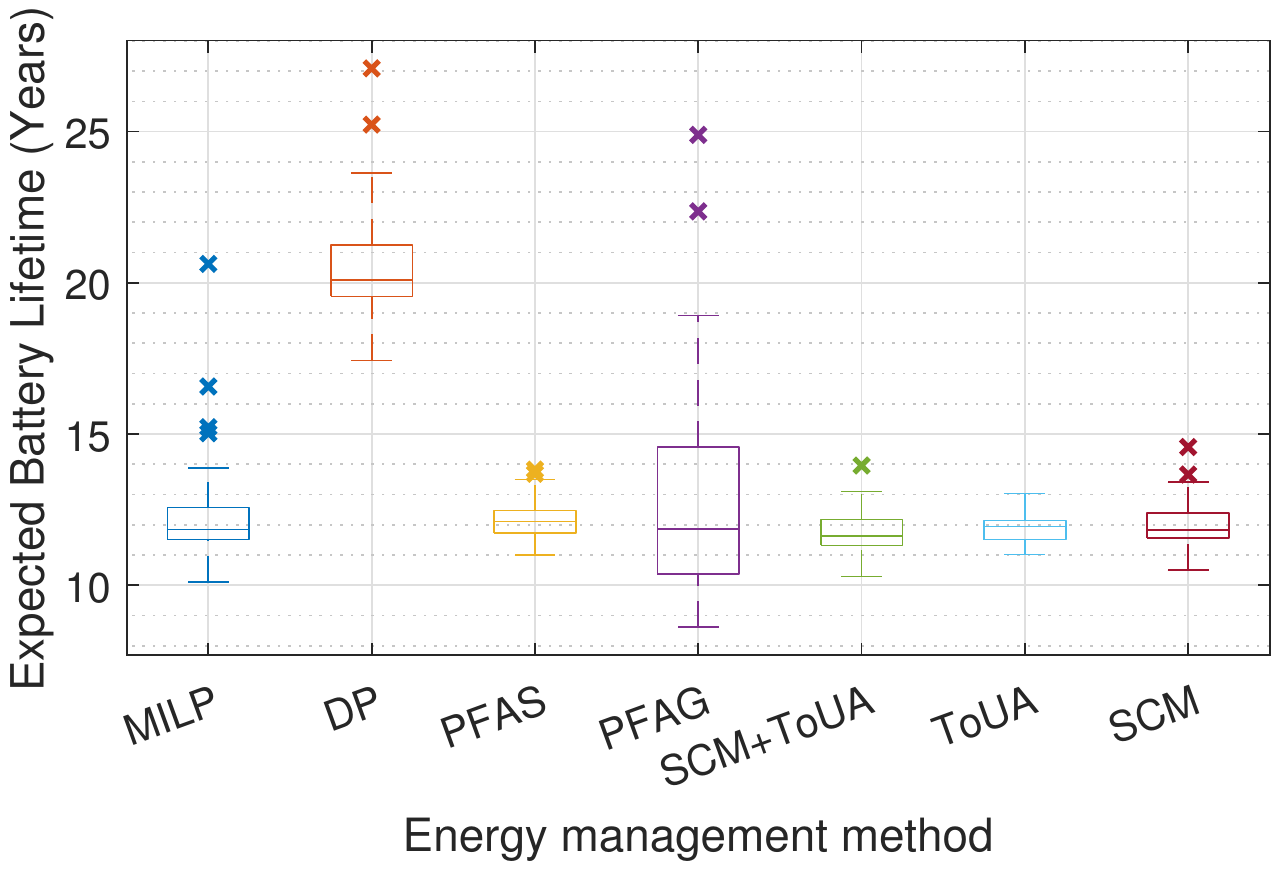} \label{figure12c}
	} \hspace{1.52em}
	\subfloat[Average cycle depth (DOC)]{%
		\includegraphics[scale = 0.65]{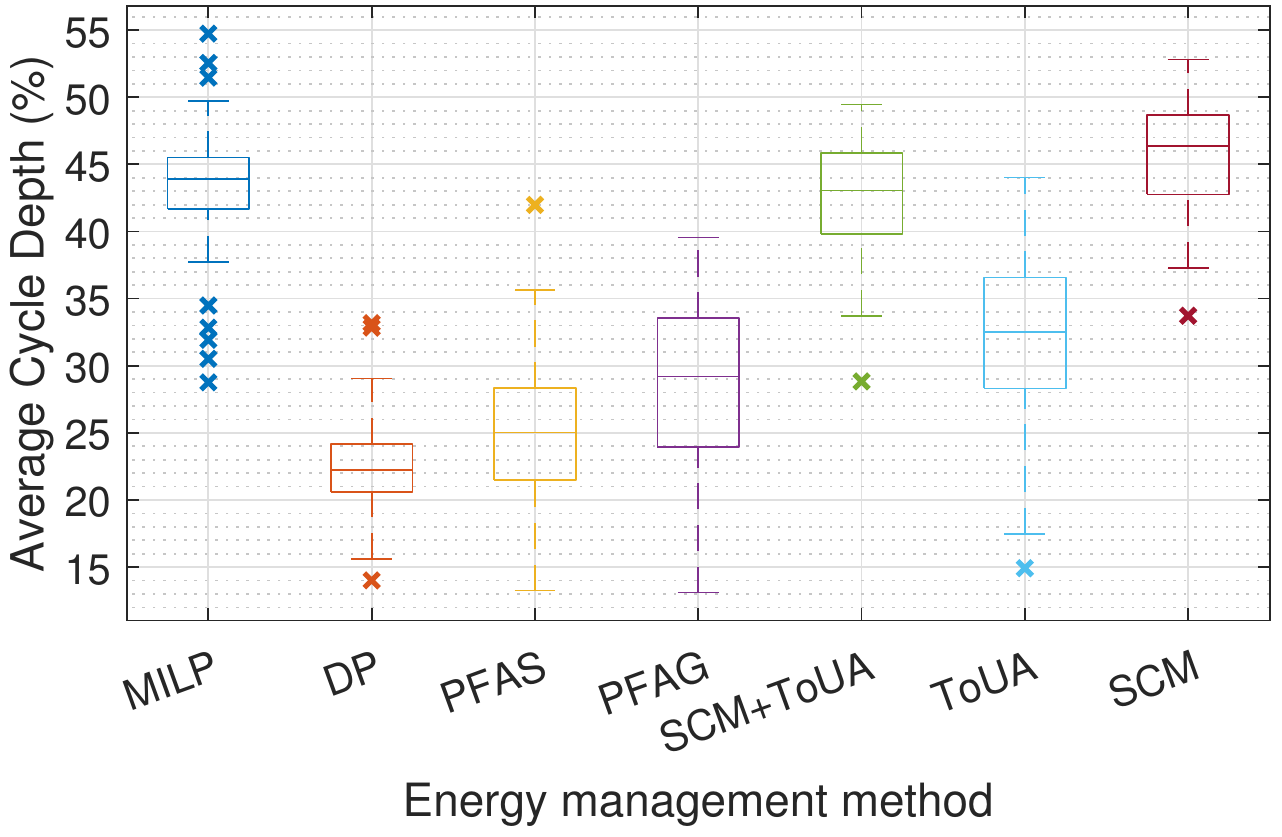} \label{figure12d}
	}
	\caption{Battery degradation results for different energy management strategies.}
	\label{figure12}
\end{figure*}

\subsection{Battery degradation results}
\par The results from the battery degradation study are given in Figure~\ref{figure12}. These include (a) the average annual full equivalent cycles, (b) the battery state of health after 20 years, (c) the expected lifetime (at 80\% SOH), and (d) the average cycle depth. 

\par As shown in Figure~\ref{figure12a}, with the optimisation approaches, the battery is subjected to a higher amount of full equivalent cycles compared to other strategies. This is due to the more frequent battery charging/discharging in response to cost minimisation. However, the average cycle depth (Figure~\ref{figure12d}) is least with DP due to the accuracy of the battery operating model. As a result, the DP strategy positively impacts the battery state of health the most. Results show a median SOH of 80\% with DP while MILP and the rule-based heuristics show similar aging performance, being based on a similar battery operating model (see Figure~\ref{figure10b}). However, ToUA performs best in this category. With ToUA, the battery is left idle during the off-peak periods, if discharging it will cause the SOC to fall below the pre-defined limit. As such, there are relatively lower loading periods with ToUA compared to MILP and the other rule-based heuristic strategies. The battery lifetime results in Figure~\ref{figure12c} follows the same pattern as the SOH results. Therefore, the expected median battery lifetime is highest with DP at 20 years, while the other strategies have values between 11--13 years. 

\par Furthermore, the machine learning approaches show similar aging results with MILP since they a trained based on the MILP battery scheduling results. Nevertheless, PFAS shows slightly better aging results compared to MILP as there are fewer full equivalent cycles. Here, the sudden jumps in battery operation with MILP are smoothed out in the neural network hence showing a smoother battery scheduling profile (see Figure~\ref{figure10c}). 

\par In summary, DP results in the best aging performance being based on a more accurate battery model. This is followed by PFAS (with the least annual full equivalent cycles), ToUA and SCM. See Table~\ref{table7} for a comparison of aging performance across the energy management strategies. It is also worth mentioning that the battery degradation performance also affects the economic viability of the PV-battery system and therefore, should be considered by the home owner when making a decision on which energy management strategy to implement.



\begin{table}[t]
	\footnotesize
	\centering
	\caption{Summary of results (lower score implies higher rank)}
	\label{table7}
\begin{tabular}{c@{\hspace{0.12cm}}c@{\hspace{0.12cm}}c@{\hspace{0.12cm}}l@{\hspace{0.12cm}}c@{\hspace{0.12cm}}l@{\hspace{0.12cm}}c@{\hspace{0.12cm}}c@{\hspace{0.08cm}}c@{\hspace{0.08cm}}}
\hline \hline
\multirow{2}{*}{\begin{tabular}[c]{@{}c@{}}Energy\\ management\\ method\end{tabular}} & \multirow{2}{*}{Speed} & \multicolumn{4}{c}{Economics} & \multirow{2}{*}{\begin{tabular}[c]{@{}c@{}}Aging\\ Perf\end{tabular}} & \multicolumn{2}{c}{Modelling} \\ 
 &  & \multicolumn{2}{c}{\begin{tabular}[c]{@{}c@{}}Perfect \\ Forecast\end{tabular}} & \multicolumn{2}{c}{\begin{tabular}[c]{@{}c@{}}Imperfect \\   Forecast\end{tabular}} &  & Complexity & Accuracy \\ \hline
SCM & 1 & \multicolumn{2}{c}{3} & \multicolumn{2}{c}{2} & 4 & 1 & 2 \\ 
ToUA & 4 & \multicolumn{2}{c}{4} & \multicolumn{2}{c}{4} & 3 & 3 & 2 \\ 
SCM+ToUA & 2 & \multicolumn{2}{c}{2} & \multicolumn{2}{c}{1} & 7 & 2 & 2 \\ 
PFAG & 6 & \multicolumn{2}{c}{7} & \multicolumn{2}{c}{6} & 6 & 6 & 3 \\ 
PFAS & 3 & \multicolumn{2}{c}{6} & \multicolumn{2}{c}{3} & 2 & 5 & 3 \\ 
DP & 7 & \multicolumn{2}{c}{5} & \multicolumn{2}{c}{7} & 1 & 7 & 1 \\ 
MILP & 5 & \multicolumn{2}{c}{1} & \multicolumn{2}{c}{5} & 5 & 4 & 2 \\ \hline
\end{tabular}
\end{table}


\section{Discussion and conclusions} 

\par In this work, we carried out a techno-economic comparative study of seven energy management strategies, including two optimisation-based approaches, two machine learning approaches and three rule-based heuristic approaches. We showed that using a more sophisticated energy management strategy may not necessarily improve the performance and economic viability of the PV-battery system because the quality of input data can be poor and because battery degradation can adversely affect the lifetime of the battery. A summary of the simulation results of these strategies in different contexts is shown in Table~\ref{table7}. As explained in Sections \ref{ecoviab} and \ref{comptime}, the results in Table~\ref{table7} may not hold true in practical implementation due to the modelling assumptions made in the different energy management strategies.

\par Our simulation results show that well-tuned heuristic strategies can give near-optimal solutions when compared to the MILP optimisation technique, assuming a linear battery model. From Table~\ref{table7}, we can conclude that SCM, which is the baseline heuristic strategy, performs close to MILP in terms of cost savings. The performance of SCM can be slightly improved when combined with time-of-use arbitrage (SCM+ToUA). 
Nevertheless, if the MILP and the rule-based heuristics are implemented on a real system, the cost savings will differ from those obtained from simulations. This is due to the assumptions made in the battery operating model. Commensurate with this, the DP results are the closest to what can be obtained in practice, given the use of a more realistic battery model. Likewise, with respect to aging performance, DP outperforms the other energy management strategies. 

\par However, if an accurate PV/demand prediction model is unavailable and there are no additional battery controllers to compensate for PV/demand prediction error, our simulation results suggests it is worthwhile implementing a simple rule-based heuristics or machine learning strategy, rather than an optimisation-based method. 
This is because with imperfect forecasts of PV and demand, the economic performance of optimisation-based approaches deteriorates since they are based on cost minimisation, explicitly modelled as the objective function.

\par Moreover, regarding to computational speed, the rule-based heuristics typically provide faster near-optimal battery schedules, compared to the other energy management strategies, assuming a linear battery model. Also, our results show that the modelling complexity is related to the computational speed and accuracy to a large extent. 
The summary results in Table~\ref{table7} indicate that the simplest strategies that are the fastest do not produce the most accurate results, indicating a trade-off between these two HEMS strategy characteristics. 
Although DP is the slowest, it is the most accurate method since it is based on a nonlinear battery and inverter model that better represents what is obtainable in practice. Nonetheless, to address its poor computational performance, it can be used in conjunction with machine learning approaches. Here, the results from DP can be used to train a PFA, to provide faster online solutions. In this way, the energy management problem can be executed with a low computational burden while still capturing the nonlinearities of battery operation. 

\par Building on the above, the use of machine learning approaches becomes more paramount on installing the PV-battery system since the customers have fewer data at this stage. Here, a plug-and-play strategy similar to PFAG will be the preferred option. With this, a new customer only needs to specify his/her consumption style and then is assigned to a particular cluster. A representative PFA representative of the cluster is then used to provide reasonably accurate, fast online solutions. 

\par Above and beyond these numerical findings, we observe that, in general, these types of results may not hold true in practical implementation due to the modelling assumptions made in the different energy management strategies. 
Thus, they should be used with caution, or ideally, verified in practise.
For future work, we plan to confirm the simulation results presented in the paper with hardware experiments using real battery systems.

\Urlmuskip=0mu plus 1mu\relax

\bibliographystyle{elsarticle-num} 



\bibliography{Economic_assessment}



\end{document}